\documentclass[12pt,twoside]{article}
\usepackage{epsfig}
\newcommand{\VolumeHeader}{}
\newcommand{\VolumeSerial}{LNS}

\newcommand{\ActivityName}{ {\normalsize {\it 
Summer School on Astroparticle Physics and Cosmology 
}}}
\newcommand{\ActivityDate}{ {\normalsize {\it
Trieste, 12--30 June 2000
}}}

\newcommand{\be}{\begin{equation}}
\newcommand{\ee}{\end{equation}}
\newcommand{\bea}{\begin{eqnarray}}
\newcommand{\eea}{\end{eqnarray}}

\newcommand{\LectureHeader}{}

\begin{document}
\pagestyle{myheadings}
\markboth{\LectureHeader}{\VolumeHeader}
\markright{\VolumeHeader}


\begin{titlepage}



\title{Finite Temperature Field Theory and Phase Transitions} 

\author{M.-P. Lombardo \thanks{lombardo@science.unitn.it}
\\[1cm]
{\normalsize
{\it Istituto Nazionale di Fisica Nucleare, 
Sezione di Padova e Gruppo Collegato di Trento,
Italy}}
\\[10cm]
{\normalsize {\it Lectures given at the: }}
\\
\ActivityName 
\\
\ActivityDate 
\\[1cm]
{\small \VolumeSerial} 
}
\date{}
\maketitle
\thispagestyle{empty}
\end{titlepage}

\baselineskip=14pt

\newpage
\thispagestyle{empty}


\begin{abstract}
These lectures review  phases and  phase transitions 
of the Standard Model, with emphasis  on those aspects which are 
amenable to a first principle study. Model calculations 
and theoretical ideas of practical applicability are discussed
as well. Contents:  1.Overview;  
2. Field Theory at Finite Temperature and Density; 
3.Critical Phenomena; 4.Electroweak Interactions at Finite Temperature;  
5. Thermodynamics of Four Fermions models; 6.The Phases of QCD;
7.QCD at Finite Temperature, $\mu_B = 0$; 
8.QCD at Finite Temperature, $\mu_B \ne 0$.
\end{abstract}

\vspace{6cm}




\newpage
\thispagestyle{empty}
\tableofcontents

\newpage
\setcounter{page}{1}

\section{Overview}

The aim of these lectures is  mostly practical: I would like
to describe the status of our understanding of  phases, and  
phase transitions of the Standard Model relevant  to cosmology 
and astrophysics, and to provide the tools to follow the current 
literature. My goal is to present results obtained from first principle
calculations, together with open problems. 
Theoretical ideas which have inspired model builders,
and  the results coming from such model calculations
will be discussed as well.

The main phenomena are the ElectroWeak finite temperature 
transition, occurring at an energy O(100 Gev)  and the
QCD finite temperature transition(s), occurring at energies O(100
Mev), within the range of current experiments at CERN and Brookhaven.
The role of a (small) baryon density in QCD will be considered as well.

Only the equilibrium theory shall be considered. The main reasons are,
firstly that equilibrium statistical mechanics is enough
to characterize many aspects of these 
phase transitions, secondly that , at variance with non-equilibrium,
equilbrium techniques are well consolidated, and at least some
interesting results are already on firm ground. The last but not
the least element in this choice is, of course, my personal expertise. 

The material is organised as follows: Section 2 reviews a few basics idea,
including the general idea of dimensional reduction, and universality.
Section 3 collects those
basic facts on phase transitions, critical phenomena, and spontaneous
symmetry breaking which are relevant to the rest.
Section 4 discusses the high temperature 
Electroweak transition : there, the perturbative
version of dimensional reduction will be seen at work, and the role
of lattice studies in elucidating crucial aspects of the Standard Model
will be emphasized. Section 5 introduces fermions,  offers
the opportunity of contrasting lattice  and 1/N expansions results,
and presents the phase diagram of four fermion models, which have
a similar chiral symmetry as  QCD. A general discussion on the phases
of QCD, and qualitative aspects of phase transitions, is given in Section 6. 
Section 7 presents the status of the
high temperature QCD studies,  discusses in detail the 
symmetry aspects of confinement and chiral transition, and the
``non--perturbative'' dimensional reduction. In Section 8 we
study QCD with a non--zero baryon density. The results for the
phase diagram of the two color model are discussed in detail, together with 
the problems and possibility for an ab  initio study of three
color QCD. In addition, it is stressed   that lattice not only makes 
a numerical ``exact'' study possible, but it is also amenable to an 
analytic treatment: the strong coupling expansion. 
We conclude by reviewing recent work which, combining universality
arguments, model calculations and lattice results,  suggests
the existence of a genuine critical point in the QCD phase diagram. 

Recent reviews and further readings include \cite{quiros} \cite{ruba96} 
\cite{zinn}  \cite{boya} \cite{thoma}. 
Closely related aspects not covered here include 
non--equilibrium phenomena \cite{bode}, recent developments 
on the high  temperature expansion in QCD \cite{iancu}, 
beyond-the-standard-model developments \cite{laru98}, 
\cite{mssm}, model calculations
in the high density phase of QCD \cite{vari}.

\newpage
\section{Equilibrium Field Theory at Finite Temperature}
The basic property of equilibrium field theory is that one
single function, $\cal Z$, the grand canonical partition function,
determines completely
the thermodynamic state of a system 
\begin{equation}
\cal Z = \cal Z (V, T, \mu)
\end{equation}
$\cal$ Z is the trace of the density matrix of the system $\hat \rho$
\begin{eqnarray}
{\cal Z} &=& Tr \hat \rho \\ 
\hat \rho &=& e^{-(H - \mu \hat N)/T}
\end{eqnarray}
$H$ is the Hamiltonian, $T$ is the temperature and $\hat N$ is any
conserved number operator. In practice, we will only concern ourselves
with the fermion (baryon) number.

Remember that $\cal Z$  determines the system's state according to:
\begin{eqnarray}
P &=& T \frac {\partial ln {\cal Z}}{\partial V} \\
N &=& T \frac {\partial ln {\cal Z}}{\partial \mu} \\
S &=& \frac {\partial T ln {\cal Z}}{\partial T} \\
E &=& -PV + TS +\mu N 
\end{eqnarray}
while physical observables $<O>$ can be computed as
\begin{equation}
<O> = Tr O \hat \rho / {\cal Z}
\end{equation}

Any of the excellent books on statistical field theory and thermodynamics
can provide a more detailed discussion of these points.
I would like to underscore, very shortly,that the  problem is to 
learn how to  represent  $\cal Z$ at non zero temperature 
and  baryon density, and to design a calculational
scheme.  In this introductory chapter we will be dealing with the first
of these tasks. Interestingly, the representation of $\cal Z$ 
which we review in  the following
naturally leads to an important theoretical suggestion, namely
dimensional reduction.

\subsection{Functional Integral Representation of $\cal Z$}
Consider the transition amplitude for returning to the original state
$\phi_a$ after a time t
\begin{equation}
<\phi_a| e^{-iHt}| \phi_a> = \int d\pi
\int_{\phi(x,0) = \phi_a(x)} ^ {\phi(x,t) = \phi_a(x)} d \phi 
e^{i\int_0^tdt \int d^3x (\pi(\vec x, t)
\frac {\partial \phi (\vec x, t)} {\partial t} - H (\pi, \phi))}
\end{equation}
Compare now the above with  expression (2)
for $\cal Z$, and make the trace explicit:
\begin{equation}
{\cal Z} = Tr e^{-\beta ( H - \mu \hat N)} =
\int d \phi_a <\phi_a | e^{-\beta (H - \mu N)} | \phi_a >
\end{equation}
We are naturally lead to the identification
\begin{equation}
\beta \equiv \frac{1}{T} \rightarrow it
\end{equation}

In Figure 1 we give a sketchy view of a field theory on
an Euclidean space. Ideally, the space directions are infinite,
while the reciprocal of the (imaginary) time extent gives the physical
temperature. 
We note that studying nonzero temperature on a lattice
\cite{lattice_rev} is straighforward:
one just taks advantage of the finite temporal extent of
the lattice, while keeping
the space directions much larger than any physical scale in the system.

\begin{figure}[htb]
\begin{center}
{\epsfig{file=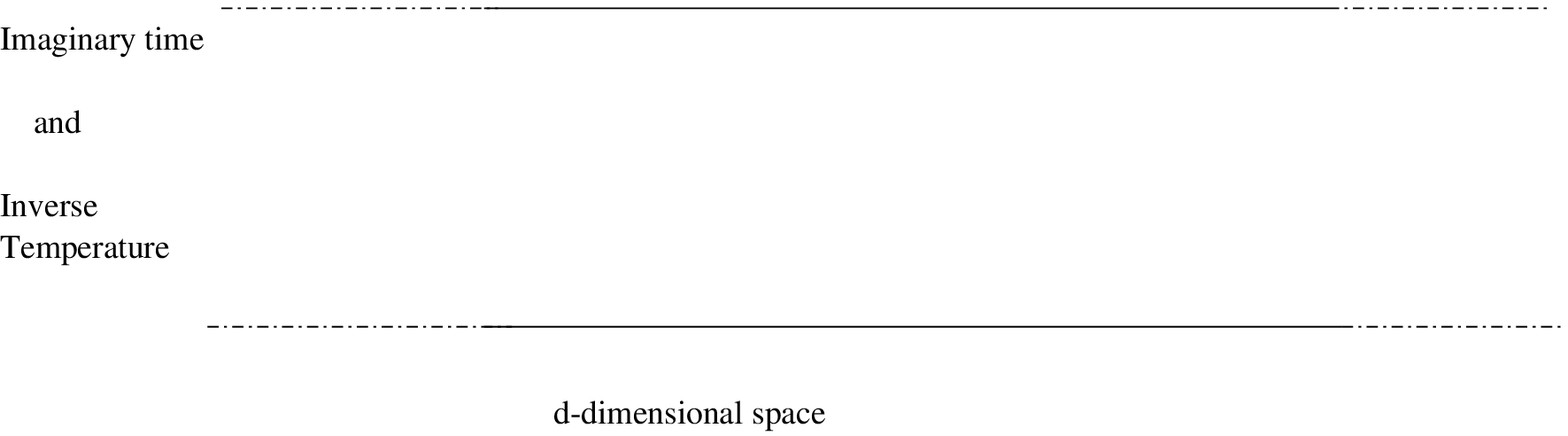, width=15 truecm}}
\caption{\footnotesize Sketchy view of the d+1 dimensional
Euclidean space. The imaginary
time is the inverse of the physical temperature of the system.}
\end{center}
\label{fig:FIG1}
\end{figure}

We need  now  to specify the boundary conditions for 
the fields. Let us introduce the integral 
$S(\phi, \psi)$ of the Lagrangian density (from now on we will
always use 1/T as the upper extremum for the time integration).
\begin{equation}
S(\phi, \psi) = \int_0^{1/T} dt \int d^d x {\cal L}(\phi, \psi)
\end{equation}
$\cal Z$ is now written as
\begin{equation} 
{\cal Z} = \int d \phi d \psi e^{-S(\phi, \psi)}
\end{equation}
To find out the boundary condition for the fields, we study the 
thermal Green functions
describing propagation from point $(\vec y, t=0)$ to point
$(\vec x, t =\tau)$
Consider the bosons first:
\begin{equation}
G_B(\vec x, \vec y; \tau, 0) = 
Tr \{ \hat \rho T_{\tau} 
[\hat \phi(\vec x, \tau)\hat \phi(\vec y, 0)]\} / {\cal Z}
\end{equation}
where $T_\tau$ is the imaginary time ordering operator:
\begin{equation}
T_\tau[\hat\phi(\tau_1)\hat\phi(\tau_2)]= 
\hat\phi(\tau_1)\hat\phi(\tau_2)\theta(\tau_1 -\tau_2) {\bf +}
\hat\phi(\tau_2)\hat\phi(\tau_1)\theta(\tau_2 - \tau_1)
\end{equation}
Use now the commuting properties of the imaginary time ordering
evolution and H:
\begin{equation}
[T_\tau, e^{-\beta H}] = 0
\end{equation}
togheter with the Heisenberg time evolution
\begin{equation}
e^{\beta H} \phi (\vec y, 0) e^{- \beta H} = \phi (\vec y, \beta)
\end{equation}
to get:
\begin{equation}
G_B(\vec x, \vec y; \tau, 0) = G_B(\vec x, \vec y; \tau, \beta)
\end{equation}
which implies
\begin{equation}
\hat \phi(\vec x, 0) = \hat \phi (\vec x, \beta)
\end{equation}
Now  the fermions. Everything proceeds exactly in the same way,
\underline{but} for a crucial difference: a minus sign in the 
imaginary time ordering coming from Fermi statistics: 
\begin{equation}
T_\tau[\hat\psi(\tau_1)\hat\psi(\tau_2)]= 
\hat\psi(\tau_1)\hat\psi(\tau_2)\theta(\tau_1 -\tau_2) {\bf -}
\hat\psi(\tau_2)\hat\psi(\tau_1)\theta(\tau_2 - \tau_1)
\end{equation}
yelding:
\begin{equation}
\hat \psi(\vec x, 0) = -\hat \psi (\vec x, \beta)
\end{equation}
Hence, fermions obey antiperiodic boundary conditions in the
time direction.

\subsection{The Idea of Dimensional Reduction}

Consider again Figure 1: it is intuitive that when the smallest
significant lenght scale
of the system $l >> 1/T$ the system becomes effectively d--dimensional.
This  observation opens the road to the possibility of a
simple description of many physical situations
(see for instance \cite{pet} for a recent review). 
It is also often
combined with the observation that the description of the system can be 
effectively `coarse grained', ignoring anything which happens on
a scale smaller than  $l$ . Again, this can be pictured in a
immediate way on a discrete lattice: the original system can be 
firstly discretized on the Euclidean d+1 dimensional space,
than dimensionally reduced to a d dimensional space, and
finally coarse grained, still in d dimensions.

In Figure \ref{fig:FIG2} we give a cartoon for this two steps procedure:
\begin{figure}[htb]
\begin{center}
{\epsfig{file=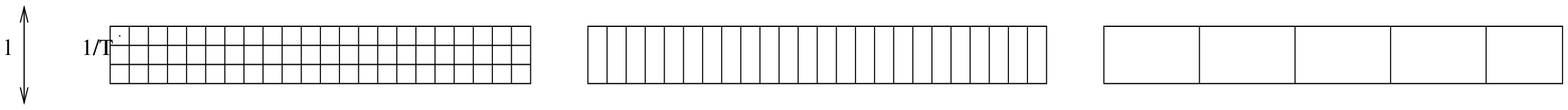, width=15 truecm}}
\caption{\footnotesize Sketchy view of the dimensional
reduction from d+1 to d dimensions(from the leftmost to the middle picture) and
subsequent coarse graining (from middle to rightmost)}
\end{center}
\label{fig:FIG2}
\end{figure}
It should be clear that, despite the simplicity and elegance of the
idea, both steps are far from trivial. In Section 4  we will
discuss the application of these idea to the Electroweak interactions
at high temperature, where the dimensional reduction will be carried
out with the help of perturbation theory. In Section 7
we will discuss dimensional reduction and universality for the
QCD high temperature transitions, where the procedure relies
on the analysis of system's symmetries. 

In general, there are two typical situations in which these idea can
be tried:

\begin{enumerate}
\item The temperature is much higher than any mass. This is
the basis for the high temperature dimensional reduction,
like, for instance, in the high T electroweak interactions.

\item The system is approaching a continuous transition:
the correlation length of the system $\xi$ is diverging.
In such situation all the physics is dominated by long wavelength
modes. Not only the system gets effectively reduced, but the
coarse graining procedure become doable. As an effect of this
procedure, systems which are very different one from another
might well be described by the same model, provided that the
long range physics is regulated by the same global symmetries: 
this is the idea of universality which provides the 
theoretical framework for
the study of the high T QCD transition \cite{piwi}.
\end{enumerate}

\subsection{Mode Expansion and Decoupling}
Fermions and bosons behave differently at finite temperature.
Consider infact the mode expansion for bosons:
\begin{equation}
\phi(x,t) = \sum_{\omega_n = 2 n \pi T} e^{i \omega_nt} \phi_n(x)
\end{equation}
where the periodic boundary conditions for the bosons have been taken
into account, and the analogous for fermions, when boundary conditions
become antiperiodic:
\begin{equation}
\psi(x,t) = \sum_{\omega_n = (2n +1) \pi T} e^{i \omega_nt} \psi_n(x)
\end{equation}
In the expression for the Action
\begin{equation}
S(\phi, \psi) = \int_0^{1/T}dt\int d^dx {\cal L}(\phi, \psi)
\end{equation}
the integral over time can then be traded with a sum over modes, and
we reach the conclusion that 
a d+1 statistical field theory at $T>0$
is equivalent to a d-dimensional theory with an 
\underline{infinite} number
of fields.

Dimensional reduction means that only one relevant field survives:
this is only possible for the zero mode of bosons.

\subsection{Finite Temperature--Summary}

\begin{itemize}
\item
The partition function $\cal Z$ 
has the intepretation of the partition
function of a statistical field theory in d+1 dimension, where the temperature
has to be identified with the reciprocal of the (imaginary) time.
\item
The fields' boundary conditions follows from the Bose
and Fermi statistics
\begin{eqnarray}
\phi(t=0,\vec x) &=& \phi(t = 1/T, \vec x) \\
\psi(t=0,\vec x) &=& -\psi(t = 1/T, \vec x) 
\end{eqnarray}
i.e. fermionic and bosonic fields obey antiperiodic and periodic
boundary conditions in time.
\item ``Dimensional reduction'', when `true' means that the system
become effectvely 3-dimensional. 
In this case only the Fourier component of each Bose field with vanishing
Matsubara frequency will contribute to the dynamics, while Fermions
would decouple.
\item The scenario above is very plausible and physically well founded,
but it is by no means a theorem. Ab initio calculations can confirm
or disprove it.
\end{itemize}
We conclude this Section by coming back to the first equation,
the one defining the partition funcion $\cal Z$. 
All we did so far is to write down 
$\cal  Z$ for a finite temperature
and density in the imaginary time formalism, 
and to examine some immediate consequences of
such a representation.

\begin{figure}[htb]
\begin{center}
{\epsfig{file=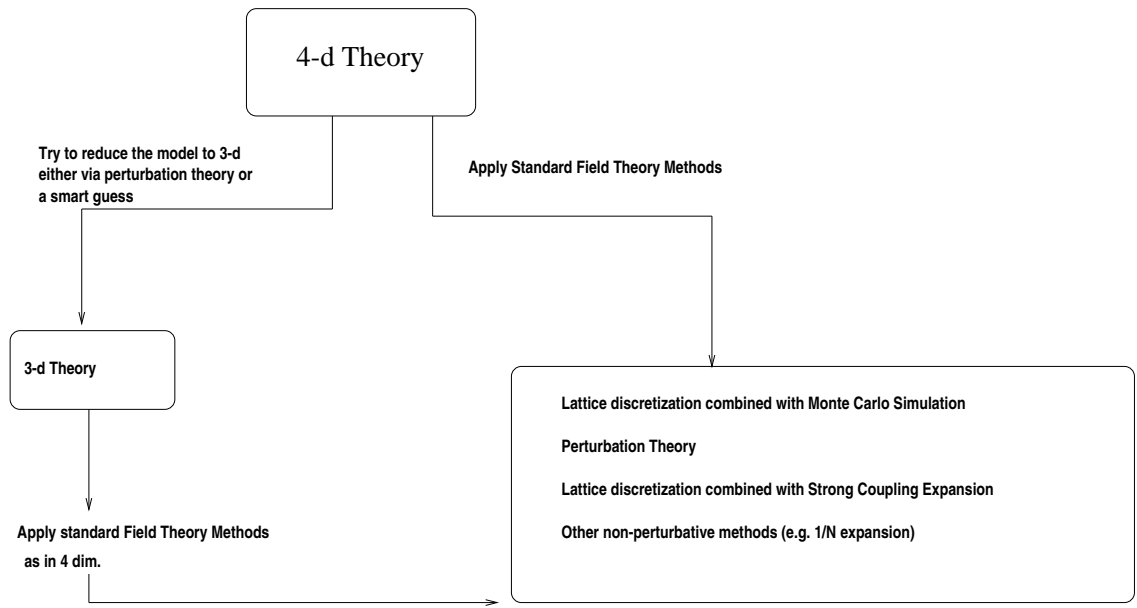, width=15 truecm}}
\caption{\footnotesize Calculational schemes for a 4d theory at a 
glance}
\end{center}
\label{fig:FIG3}
\end{figure}

In Figure \ref{fig:FIG3}  we give  a sketch which outlines some of 
the various possibilities to calculate {\cal Z} for
a four dimensional model at finite T formulated in the
imaginary time formalism. 

Two  remarks are in order.
Firstly,  the optimal strategy is model dependent and there
is no general rule which can tell us a priori which is the best way
to attach a problem. In the following  we will see the various
techiques at work in different models and situations.

Secondly,  there are idea which can be of general applicability. 
This is especially true for  universality, for which a general
framework is given by the symmetry analysis, breaking patterns and 
theory of critical phenomena. We then devote the next Section to a 
survey of this subject.

\section{Critical Phenomena}
The modern era of critical phenomena begun about fifty years ago, when
it was appreciated that very different fluids might well display
{\em universal} behaviour: the cohexistence curve, namely the plot
of the reduced temperature $T/T_c$  versus the reduced density 
$\rho/\rho_c$ is the same! This observation has triggered 
important theoretical developments, in particular it has emerged that
many phase transitions are associated with a spontaneous breakdown of
some symmetry of the Hamiltonian, and that
these symmetry changes can be triggered by temperature
\cite{doja}. 
In this Section we briefly review main tools and idea used in this field.

\subsection{The Equation of State and the Critical Exponents}
In a broad sense, the Equation of State is a relationship among
different physical quantities characterising the critical
state of a system, which only depend on a few
parameters. For instance for fluid systems it was soon realised that
a general relationship describing the cohexistence curve is:
\begin{equation}
(\rho - \rho_c) = A \bigg( \frac{T - T_c}{T_c} \bigg ) ^\beta 
\end{equation}
where $T_c, \rho_c, A$ depend on the specific fluid under consideration, 
while the functional form above with $\beta \simeq 3 $ 
holds true for all the fluids which have been studied.

It came then natural to apply the same idea to the critical phenomena
in magnetic systems, identifying, in a completely euristic way, the
zero field magnetization $M$ with the density difference:
\begin{equation}
M = A \bigg( \frac{T-T_c}{T_c}\bigg)^\beta
\end{equation}
which can be easily generalised to include an external magnetic
field h
\begin{equation}
\frac{h}{M^\delta} = f \bigg(\frac{T - T_c}{M^{1/\beta}} \bigg)
\label{eqs:eos}
\end{equation}
The above equation is the well known equation of state for magnetic
systems, in one of its guises. For $h=0$ we recover eq. (28) , together
with the definition of the exponent $\beta$. For $T = T_c$ we have
\begin{equation}
M = h ^{1/\delta}
\end{equation}
and we recognize that $\delta$ is the exponent describing the
system's response at criticality. It is really remarkable that
these definitions can be combined in just one relationship,
the equation of state eq.(29).
In short summary, then
\begin{itemize}
\item The Equation of State contains the usual definition
of critical exponents (which is obtained by 
setting $h = 0$ or $T = T_c$)
\item It gives information  on the critical behaviour (i.e.
the exponent $\delta$) 
{\em also by working at} $T \ne T_c$
\item It gives information on spontaneous magnetization 
(i.e. the exponent $\beta$) {\em also
when there is an external magnetic field} h
\end{itemize}
The function f, in general, is unknown (it is possible to
give some of its properties on general grounds \cite{kkl}). 
It is useful  to 
consider a first order expansion of the Equation of State
\begin{equation}
h = a M^\delta + b (T-T_c) M^{\delta  - 1/ \beta}
\label{eqs:eos_e}
\end{equation}
which is more easily amenable to a direct comparison
with data. Clearly, the range of applicability of
eq.\ref{eqs:eos_e} is smaller than the one of 
eq.\ref{eqs:eos}.

The idea outlined above can be applied in a broad range of context,
including particle physics. We will discuss very many examples
of this in the following, while in the following Table, as a first example,
we summarize a `dictionary' between magnetic systems, bosonic systems, 
and fermionic ones, where
the order parameter is a composite fermion-antifermion condensate:
\newpage
\begin{tabbing}
Response Function \= M (magnetization)uuuu\= ttttttttt$<\sigma>(m)$uuu \= uuuu
$<\bar\psi\psi>(m)$ 
\kill \\
 \> {\bf Magnets} \> {\bf Bosons} \> {\bf Fermions} \\
External  Field \> h (magnetic field) \> $m$ (bare mass) \> $m$ (bare mass) \\
Response Function \> M (magnetization)\> $<\sigma>(m)$\> $<\bar\psi\psi>(m)$ \\
Order Parameter\> $M_0=M(h=0)$  \> $<\sigma> (m=0)$ \>$<\bar \psi \psi>(m=0)$\\
\end{tabbing}

The systems above can be described in an unified way via the Equation
of State we have discussed. Systems whose critical behaviour
is characterised by the same exponents  are said
to belong to the same universality class.  
This circumstance  is of course highly non trivial
and it is rooted in the symmetries of the systems: systems whose critical
behaviour is governed by the same global symmetries are in the
same universality class. Strictly speaking this concept, as well as
the equation of state itself,is limited to continuos transitions.
Continuous transition have real divergencies,
 where the correlation length grows so large that it is possible to
ignore all of the short range physics, but the one associated with the
bosonic zero modes of the system. When this happens details of the
dynamics do not matter any more, and the pattern of symmetries,
their spontaneous breaking and realisation,  drive
the system's behaviour.

\subsection{The Effective Potential}

The Equation of State provides a  macroscopic description
of the order parameter of the system close to criticality.
It is very useful and interesting to consider
an intermediate level of description (in condensed matter/statistical
physics it might be called mesoscopic) where the
behaviour of the order parameter is inferred from its probability 
distribution (or for any distribution related with it). 
In turn, such probability distribution can/should
be derived from the exact dynamics, with the help of a symmetries'
analysis.

\begin{figure}[htb]
\begin{center}
{\epsfig{file=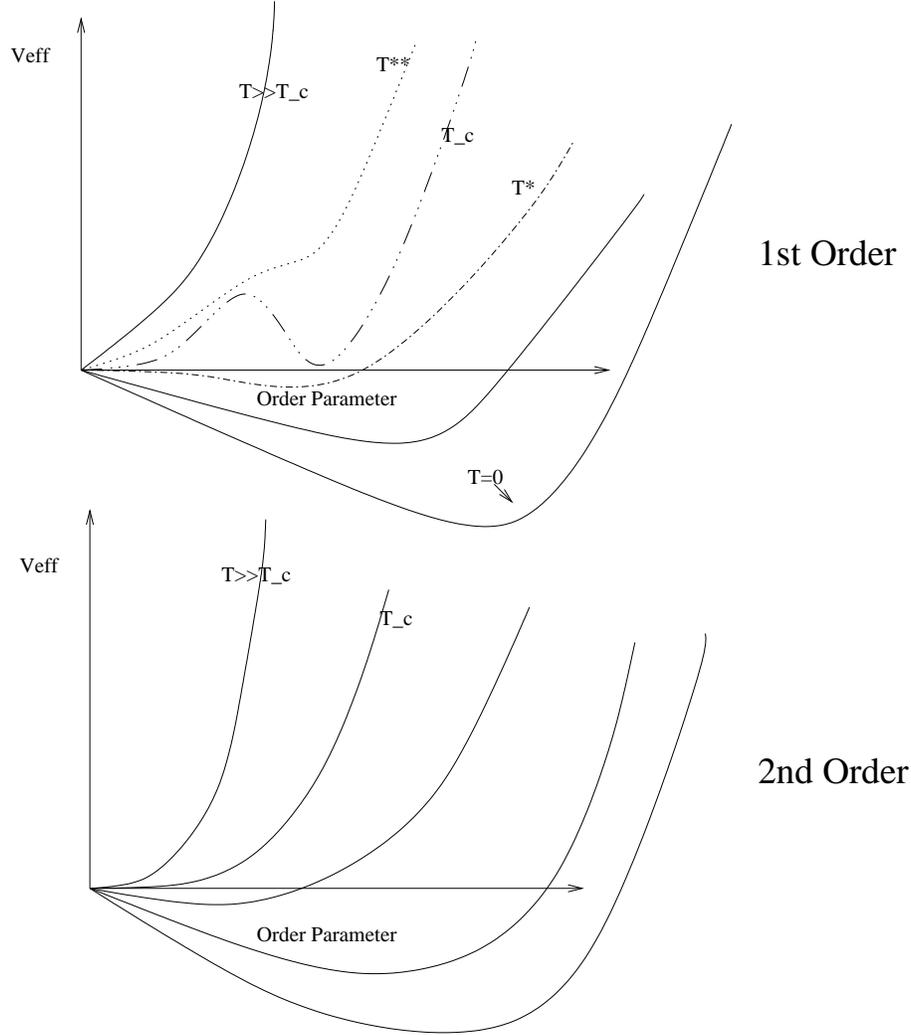, width=12 truecm}}
\caption{\footnotesize The shape of the effective potential for
the order parameter for first order, discontinuous 
(upper diagram) and continous transitions.}
\end{center}
\label{fig:FIG4}
\end{figure}

Main idea goes back to Landau, and is the following. One is supposed
either to guess or to derive a function $V_{eff}$ of the order
parameter and external fields which describes the state of the system.
We draw in Figure \ref{fig:FIG4} the familiar plots showing
$V_{eff}$ as a function of the order parameter for
several value of the external field (the temperature, for instance).
The minimum of $V_{eff}$ defines the most likely value of the order
parameter, and so these plots determine the order parameter as a function
of temperature. Also, for any give temperature we 
read off the plots wheater the system is in a pure state 
(just one minimum), or in a mixed phase (two non equivalent minima), 
which is only possible  for a first order transition.
In the upper plot we show the behaviour leading to  a 
first order transition.
For low temperature the system is in a phase 
characterised by a nonzero 
value of the order parameter. At $T = T^*$ 
(spinodal point) we have the first occurrence
of a secondary minimum corresponding to a zero value of the order
parameter, i.e. the onset of a mixed phase.
For $T = T_c$ the two minima are equal, i.e. $T_c$ is the critical point. 
Beyond $T_c$ the minimum  at zero
is the dominant one. At $T = T^{**}$ the secondary minimum (for a nonzero
value of the order parameter) disappears, and the mixed phase ends.
In the lower diagram it is shown an analogous plot, but for a continuous 
(second order) transition. We note that a second order transition can be 
seen as a limiting case of a first order transition for $T^*, 
T^{**} \to T_c$. 

\subsection{A Case Study }

The idea discussed above can be nicely elucidated by the Potts model.
There, one can find all possible examples of phase transitions,
from continuos to discontinuous, and different intensities.

Consider the Action for the q-state two dimensional Potts model:
\begin{equation}
S = - \beta \sum \delta_{\sigma_i \sigma_j}
\end{equation}
The spin variables $\sigma$ can be in any of the $q$-possible states.
The sum runs over nearest neighboors, and, for small $\beta$
clusters of equal $\sigma$ are preferred. When $\beta$ increases
there is a transition to a disordered state, for $\beta = \beta_c$.

The model has been analytically solved in a mean field approximation, where
one can calculate $\beta_c$ and the latent heath for any $q$.
The latent heat turn our to be zero for $q \le 4$, corresponding  to a second 
order transition, and positive, and increasing with $q$, for q larger than
4. For q very close to 4, the behaviour is {\em weakly first order}
or {nearly second order}, in the sense that the first order
transition behaves like a second order one, near, but not too
near, the critical point. For instance, the inverse correlation
length should obey the law
\begin{equation}
\xi ^{-1} = A(\beta^* - \beta)^{\nu*}
\end{equation}
where $\beta^*$ (to be identified with the $T^*$ of
the discussion above) and $\nu*$ can be tought of as the coupling and critical
exponent of the virtual second order transition point in the metastable
region.

This behaviour have indeed been observed in numerical simulations,
which can be described using the associate Ginzburg Landau model
\cite{ruiz}.
For instance the transition in the seven state Potts has a much smaller
latent heat than the one with ten states, while the pseudocritical
point $\beta^*$ is much closer to $\beta_c$ for $T=T_c$ for seven states
than for ten.

One might think that the concept of a weak first order transition is
a rather academic one. Indeed, it contrasts a bit with the familiar notion
of an abrupt discontinuous transition, without precursor effects.
It is then amusing to notice that a weak first order behaviour has
been indeed observed not only in liquid cristals--the original
de Gennes discussion on weak first order transitions-- but also in 
the deconfining transition of Yang Mills model with SU(3) color group-a 
transition which will be discussed at length in Section 7 --
The strength of a first order transition is also an issue in the
discussion of the electroweak transition \cite {gleikolb}, as we
will review in the next Section.

\subsection{Dynamical Symmetry Breaking}

It is a fact that often the symmetries of the Hamiltonian
are not realised in nature: for instance the complete translational 
invariance of a many body system is broken when the system
becomes a cristal, or the chiral symmetry of the QCD
Lagrangean is broken when  quarks and antidiquarks condense.
The realisation, or lack thereof, of the symmetries of the Hamiltonian,
i.e. the existence of different phases in a system, 
depends on the thermodynamic conditions. Hence, many phase transitions
and critical phenomena are associated with a symmetry
breaking/restoration\cite{vile}.

One simple model to study this is the Goldstone model
\begin{equation}
L = (\partial_{\mu}  \phi ^\dagger)(\partial^\mu \phi) - V(\phi)
\end{equation}
where $\phi$ is a complex scalar field and the potential
$V(\phi)$ is given by
\begin{equation}
V(\phi) = 1/4 \lambda (\phi^\dagger \phi - \eta^2)^2
\end{equation}

The Lagrangian is obviously invariant under $U(1)$ global
phase transformation of the field $\phi$. The only
value of the field which realises such invariace is zero.

For positive $\lambda$ and $\eta^2$ $V(\phi)$
has the well known Mexican hat shape: there is an infinity
of minima of the effective potential for a nonzero
value of the field. Once one particular minimum is selected, 
$\phi \ne 0$ and the symmetry is broken. 
 
If $\eta$ is tuned towards zero the global minima disappear,
and the symmetry is restored: the only global
minimum is the one corresponding to a zero value for the
field.

It is worth noticing that simple models 
have also been used as toy models in various situations in
particle, astroparticle or condensed matter physics:
the nonrelativistic version of the Abelian Higgs model is identical
to the Ginzburg-Landau model of a superconductor with $\phi$ 
representing the Coopper pair wave function, the Goldstone
model describes the Bose condensation in superfluids,  the 
$\sigma$ model can be used to model the QCD high T phase transition,
$\sigma$ describing the composite $<\bar \psi \psi>$ condensate,
the three state Potts model models the deconfining transition
in three color Yang--Mills, the spins representing the Polyakov
loop,  the  non--Abelian Higgs model  is used to study 
the electroweak  transition, $\phi$ being in this case the Higgs field, 
etc. etc.

These toy models are amenable to approximate, mean field studies
but to get real predictions one needs to study the full
field theory from first principles. It is always very interesting
to try to make contact between the exact results and those derived
from simple models and analysis, so to assess the role of the quantum
fluctuations. We will see  examples of these points in the 
rest of these lectures, beginning with the electroweak transition 
in the next Section.

\section{Electroweak Interactions at Finite Temperature}

We want to know the fate of the spontaneously broken gauge symmetry of 
the electroweak interaction at high temperature. This depends 
on the value of the Higgs mass $m_H$, and cannot be solved within
perturbation theory. This question has been sucessfully addressed 
by Laine,Kajantie,Rummukainen,Shaposhnikov in a series of
papers \cite{kalarusha} \cite{theta0} \cite{eff3d} \cite {laru98},
and references therein. These authors have been using
an admixture of perturbative dimensional reduction and lattice 
calculations, which eventually
gives the phase diagram of the Electroweak sector of
the Standard model in the $m_H,T$ plane.
A classical reference on the ElectroWeak transition
and its phenomenological implications is \cite{ruba96}. 

The essence of the model can be found without inclusion of fermions.
This is good news as, despite substantial progress in the formulation
of chiral fermions on the lattice
\cite{chiral_news}, practical calculations, at least in four dimensions, 
are still rather difficult. 
In addition to disregarding fermions, the effects of the
U(1) group shall be neglected as well ($sin^2 \theta \simeq 0$), since
its effects have been found to be small \cite{theta0}.

All in all, the Lagrangian we study is
\begin{equation}
L = 1/4 F^a_{\mu \nu}F^a_{\mu \nu} + (D_\mu \phi)^\dagger(D_\mu \phi)
- m^2 \phi^\dagger \phi + \lambda (\phi^\dagger \phi)^2
\end{equation}
where $\phi$ is the Higgs doublet.

\subsection{Perturbative Analysis}
We begin by briefly reviewing the perturbative results,
see \cite{gleikolb} \cite{cline} \cite{ruba96} for details.
Let us then build an effective potential from the electroweak Action by 
approximating the field $\hat \phi$ by a constant value $\phi$, i.e.
neglecting the fluctuations:
\begin{equation}
VT V_{eff} (\phi) = \int dt d^3 x  S (\hat \phi) \delta (\hat\phi - \phi)
\end{equation}
The zero temperature tree level results reads is:
\begin{equation}
V_{eff}(\phi, T=0) = - m^2/2 \phi^2 + \lambda/4 \phi^4
\end{equation}
(we can just consider the discrete $\phi \rightarrow - \phi$ symmetry
which is enough to describe the behaviour of the order parameter).

The $T > 0$ one loop corrections basically are 
the effects of a non-interacting bose-fermi gas with frequencies
depending on $\phi$ (the -/+; +/- signs are for bosons/fermions):
\begin{eqnarray}
V_{eff}^{1-loop} (\phi, T) &=& T \sum_i -/+ \int \frac{d^3p}{(2p)^3} ln
\int 1 +/- e^{- (p^2 + \omega^2(\phi))^{1/2}/ T} =\\
&& \frac{1}{2} \gamma \phi^2 T^2  \\
&&  - \frac{1}{3} \alpha \phi^3 T
\end{eqnarray}
where both bosons and fermions contribute to the term in
$\phi^2 T^2$ , and the term in $\phi^3 T$ is from bosons alone.
This yields:
\begin{eqnarray}
V_{eff} (\phi, T) &=&
 - \frac{m^2}{ 2} \phi ^2 + \frac{1}{2} \gamma T^2 - 
\frac{1}{3} \alpha T \phi ^3 + 
\frac{1}{4} \lambda \phi ^4 \\
&=& \frac{1}{2} \gamma (T^2 - T^{*2}) \phi ^2 - 
\frac{1}{3} \alpha T \phi^3 +
\frac{1}{4} \lambda \phi ^4.
\end{eqnarray}
We have defined $T^{*2} = m^2/ \gamma$ and  we will recognize
that $T^*$ is the pseudocritical point.  

At T = 0 we have then the typical case of spontaneous symmetry breaking,
with $\phi^2 = 2 m^2 / \lambda$. 
At `large' T  the  positive quadratic term $\gamma T^2$ dominates and 
symmetry is restored.

The cubic term drives the transition first order. 
The onset for metastability, i.e. of the mixed phase, 
coincides with the first appearance of a secondary miminum.
It is then given by an inflexion point with zero slope:
\begin{eqnarray}
\frac {\partial^2 V} { \partial^2 \phi } &=& 0 \\
\frac {\partial V} { \partial \phi} &=& 0
\end{eqnarray}
The solution is, as anticipated, $T=T^*$.

The critical temperature, from the equal minimum condition
\begin{eqnarray}
V(\phi_1) = V (\phi_2)
\end{eqnarray}
is 
\begin{equation}
T_c = T^*/ \bigg( 1 - \frac{2 \alpha^2}{ 9 \lambda \gamma}\bigg)^{1/2} > T^*
\end{equation}
and it depends linearly on the Higgs mass:
\begin{equation}
T_c = m_H / \bigg( 2 \gamma - \frac{4 \alpha^2}{9 \lambda}\bigg)^{1/2}
\end{equation}

As discussed in Section 3.3,
the intensity of the transition is given either by the magnitude of the jump
at $T_c$, $2 \alpha/ 3 \lambda$, and by
the distance between $T_c$ and $T^*$
\begin{equation}
T_c - T^* = T^*
\big(1/ \big( 1 - \frac{2 \alpha^2}{ 9 \lambda \gamma}\big)^{1/2} -1\big)
\end{equation}

In conclusion the perturbative analysis of the electroweak
model at high temperature predicts a first order transition. 
The transition weakens when $\alpha \to 0$ or $\gamma \to \infty$
or $\lambda \to \infty$, eventually becoming second order when
the strength goes to zero and $T_c = T^*$ (cfr. again 
the behaviour of the Potts model \cite{ruiz}). 
Note that the quantum fluctuations
are directly responsible for the weakening of the first order transition
\cite{gleikolb}. The complete inclusion of quantum fluctuations,
which we are going to discuss, will cause the disappearance of
the phase transition at large Higgs masses.

\subsection{Four Dimensional Lattice Study}
We know that in the electroweak sector of the Standard Model perturbation 
theory works well at zero temperature. As was pointed out in ref. 
\cite{kalarusha} this is not true in general. It has been recognized
that there are several scales in the problem. The smallest, which
serves as a perturbative expansion parameter, is proportional 
to $ g^2 T/m$, where
$m$ is any dynamically generated mass.
Now, dynamically generated masses will be zero in the symmetric
phase, hence perturbation theory will not be applicable there.
And also in the broken phase, according to the intensity of the
transition, perturbation theory might fail: when the 
transition is weakly first order, the mass approaches zero in 
the broken phase  and perturbation
theory is no longer reliable.

The exact approach uses ab initio lattice calculations on a four
dimensional grid. These calculations are extremely expensive.
The lattice spacing, which serves as an ultraviolet cutoff, must
be smaller that the reciprocal of any physical energy scale:
typically, at very high temperature, it must be smaller than $1/T$.
On the other hand, the lattice size in space direction must be large enough
to accomodate the lighter modes which are proportional to 
$ 1/{g^2 T}$, where $g$ is the 
lattice coupling. As the continuum limit is found for $g \to 0$,
also the requirement on size is quite severe.

In their numerical studies 
(see \cite{laru98} for a review) 
the authors selected a few significant values
for the Higgs mass, and studied the behaviour as a function of the
temperature. The critical behaviour is assessed using the tools
developed in statistical mechanics, finite size scaling
of the susceptibilities, search for two states signal, etc.
This analysis led to these results: for small values of the Higgs mass
the 4-d simulations agree with perturbation theory, as expected.
The first order line, which in perturbation theory weakens
 without disappering,
{\em ends} in reality, for an Higgs mass $\simeq 80 Gev$. The 
discovery of such an endpoint 
\cite{kalarusha} has been a major result, as it shows
that there is no sharp distinction between the phases of
the electroweak sector. The properties of this endpoint have
been investigated in detail, confirming its existence, and placing 
it in the  universality class of the 3d Ising model (a rather
general results for endpoints).

\subsection{Three Dimensional Effective Analysis}

The main idea of the three dimensional effective analysis 
\cite{eff3d} is to
use perturbation theory to eliminate the heavy modes, in the general spirit
of dimensional reduction: this makes it definitevly more easy to handle
than the four dimensional theory. In the first place, three dimensions
are of course cheaper than four. Secondly, and perhaps even most 
important, eliminating the heavy modes makes less severe the demand on
lattice spacing. In conclusion, one can simulate the three dimensional
theory on a much coarser lattice than the four dimensional one.
The advantage over the four dimensional simulations is clear.

On the other hand, the three dimensional approach is much better than 
perturbation theory: we have noticed that perturbation theory cannot
handle the light modes which might appear close to a (weak first order
or second order) phase transition. Instead, the three dimensional model
retain the light modes of the full model, only the heavy modes are integrated
out.

But which model one should simulate? The model has the same functional
form as the original one, i.e. it retains all of the particle content
and symmetries:

\begin{equation}
L = \frac{1}{4} F^a_{\mu \nu}F^a_{\mu \nu} + (D_\mu \phi)^\dagger(D_\mu \phi)
- m_3^2 \phi^ \phi + \lambda_3 (\phi^\dagger \phi)^2
\end{equation}
where the subscrit `3' reminds us that we are working in three dimensions.
The rule developed in \cite{eff3d} yields to an expression of
$(g_3, m_3, \lambda_3)$ in terms of the parameters of the 
four dimensional model. This is of course not trivial : for instance, 
in the three dimensional model there is no `obvious' temperature dependence,
as the fourth dimension has disappeared. All the temperature dependence is
`buried' in the parameters of the dimensionally reduced model.

The general arguments leading to the formulation,and applicability
or the three dimensional effective  theory are very convincing, but, 
as the authors 
pointed out,  cross checks with the full four dimensional 
simulations are needed.

\subsection{The Phase Diagram of the EW Sector of the
Standard Model}

\begin{figure}[htb]
\begin{center}
{\epsfig{file=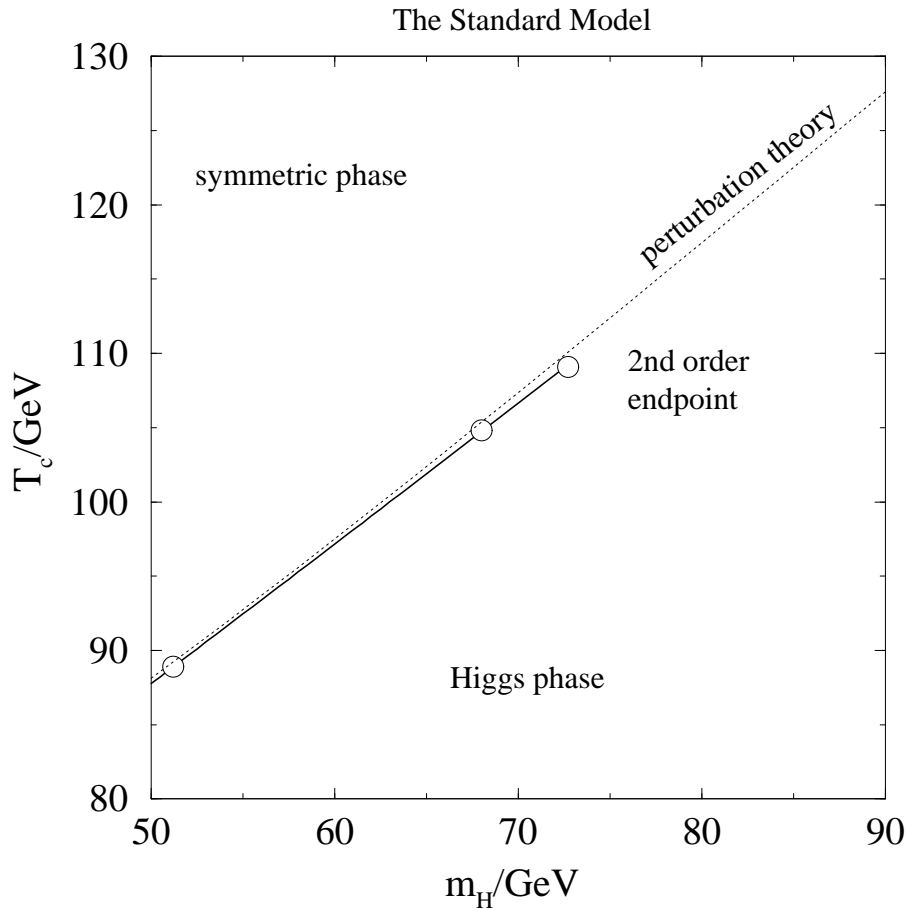, width=12 truecm}}
\caption{\footnotesize The phase diagram of the Electroweak Sector
of the standard model in the temperature, Higgs mass plane,
taken from Laine-Rummukainen \cite{laru98}}
\end{center}
\end{figure}

In Figure 5, 
reproduced from the Laine-Rummukainen review of 1998
ref. \cite {laru98}, we see
the phase diagram of the electroweak sector of the standard model,
built by use and cross checks of the results obtained 
by the methods described above.

For small values of the Higgs mass
$m_H < m_W$ the transition is strongly first order,
and weakens (in the sense discussed in Section 3 above) when $m_H$ increases.
These results can be obtained in perturbation theory. 
For large Higgs mass ($m_H > m_W$) perturbation theory is no 
longer applicable, and a
combination of lattice calculations of the four dimensional theory and
of the (perturbatively reduced) three dimensional theory  gives
fully satisfactory results.

The endpoint was discovered in \cite{kalarusha}.
The nature and location of
the second order endpoint, the most important result, has been
actively investigated since then. From a phenomenological point
of view, it is important to notice the transition ends at a value
smaller than realistic Higgs masses.

The dotted line is the
perturbative result reviewed in Section 4.1 : the limitation
of the perturbative analysis is evident from the plot.

The fact that at small mass there is no phase transition, i.e.
the discovery of the end point   shows that  there is no sharp distinction
between the symmetric and broken phase of
the electroweak sector of the standard model, much like between liquid water
and vapor. For instance, in principle 
massive W bosons in the broken phase cannot be 
distinguished from massive, composite objects in the symmetric phase. 

Current research investigates the fate of these results in various 
supersymmetric extensions of the Standard Model
\cite{laru98} \cite{mssm}. Of particular
interest is the possibility that the line of first order phase
transition continues up to realistic Higgs masses.

\section{Thermodynamics of Four Fermions Models}

Models discussed so far only included bosons. As a warmup for
the study of QuantumChromodynamics we discusse here purely
fermionic models. Besides their illustrative value, such
models  can offer some insight into real QCD.
In a nutshell,  the idea is that the interaction between the massive quark 
can be described by a short range (in effect, contact) four-fermion
interaction --  
to be specific, let us just remember the effective Lagrangian
with the t'Hooft instanton-mediated interaction:
\begin{equation}
{\cal L_I} = K_I [(\bar \psi \psi)^2 + 
(\bar \psi i \gamma_5 \vec \tau \psi)^2 - ( \bar \psi \vec \tau \psi)^2
- (\bar \psi \gamma_5 \psi)^2]
\label{eqs:thooft} 
\end{equation}
In the standard treatment of this effective Lagrangean spontaneous
mass generation is imposed by fiat, and the resulting Lagrangean
contains a mass term
\begin{equation}
{\cal L}_{eff} = \bar \psi (i \gamma^\mu D_\mu - m_\star +
\mu \gamma_0) \psi - 1/4 F_{\mu \nu} F^{\mu \nu} + {\cal L}_{int}
\end{equation}
which explicitely breaks the global chiral symmetry of 
(\ref{eqs:thooft}).

More generally, and perhaps more interestingly, 
fermionic models with a contact four fermion
interaction display, in a certain range of parameters,
spontaneous symmetry breaking and {\em dynamical} mass generation.
They are thus a playground to study these phenomena in ordinary conditions,
together with their fate at high temperature and density.
This, besides being helpful to study QCD, it is also important
from a more general point of view. A caveat is however
in order before continuing:
these models cannot describe confinement.

This section is  devoted to the study of the thermodynamics
of four fermion models with exact global chiral symmetry, see
\cite {kli} \cite{rose} \cite{hands_kogut} 
\cite {kos} for early studies, reviews, details
and recent developments. 
In a thermodynamics context,
an important observation is that purely fermionic systems do not
have zero modes, because of their antiperiodic boundary conditions.
Hence, they are not amenable to direct dimensional reduction.
We will see how the bosonised form of the models helps in this,
and other respects. In particular, the bosonised form is amenable both
to a simple mean field (large N) analysis,as well as to an
exact lattice study.Let us also mention right
at the onset, that these models (as opposed to QCD) can be studied
exactly be means of lattice calculations also at nonzero density.

\begin{figure}[htb]
\begin{center}
{\epsfig{file=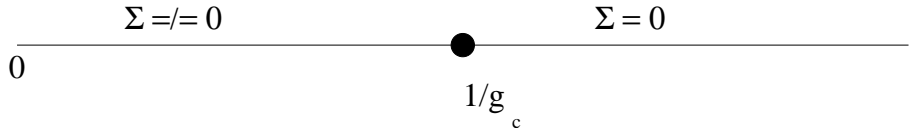, width=12 truecm}}
\caption{\footnotesize The phase diagram of the 3d Gross--Neveu model
at $T=\mu = 0$. In the large coupling region the model has spontaneous
symmetry breaking, and in that region is interesting to study the effects
of temperature and chemical potential, which might have some similarity to
QCD. The symmetric region is not 
relevant for thermodynamics.}
\end{center}
\label{fig:FIG5}
\end{figure}
The be specific, from now on we will concentrate on the 3dimensional
Gross Neveu model, described by the Lagrangian
\begin{equation}
L = \bar \psi (\slash\partial + m) \psi - g^2/N_f[
(\bar \psi \psi)^2 - ( \bar \psi \gamma_5 \psi)^2 ]
\end{equation}
The model is invariant under the   chiral symmetries $Z_2$ and $U(1)$ 
\begin{eqnarray} 
\psi_i & \rightarrow & e^{i \alpha \gamma_5} \psi_i \\
\bar \psi_i & \rightarrow & \bar \psi_i e^{i \alpha \gamma_5}
\end{eqnarray}
The main properties of interest are:
\begin{itemize}
\item For large coupling, at $T=\mu=0$, it displays spontaneous symmetry
breaking, Goldstone mechanism and dynamical mass generation
\item It has a rich meson spectrum, including a `baryon' (infact,
the fermion
\item It has a non--trivial (interacting) continuum limit
\item It is amenable to an exact ab initio lattice study
\end{itemize}
As mentioned above, its analytic and numerical study is helped 
by auxiliary bosonic fields $\sigma, \vec \pi$. $\sigma$
and $\vec \pi$  can be introduced by adding to $\cal L$
the irrelevant term $(\bar \psi \psi + \sigma / g^2)^2 
+ (\bar \psi \gamma_5 \psi + i \pi \gamma_5 / g^2)^2$.
$\cal L$ then becomes(we  set $m=0$):
\begin{equation}
{\cal L} = \bar \psi (\slash \partial +  \sigma + i \pi \gamma_5) \psi
+ N_f / 2g^2 (\sigma^2 + \pi^2)
\end{equation}
The continuous symmetry of the system is now reflected also by the rotation
in the $(\sigma, \vec \pi)$ chiral sphere.

A simple mean field analysis reveals the phase structure of the
model at $T=\mu=0$ which we skecth in Fig. \ref{fig:FIG5}.
This is obtained by solving for the expectation value of the dynamical 
fermion mass $\Sigma \equiv < \sigma>$ via the gap equation 
\begin{equation}
\Sigma = g^2 \int_p tr \frac{1}{i \slash p + \Sigma}
\end{equation}
yelding
\begin{equation}
\frac {1}{g^2} = - \int \frac {d^3 p}{p^2 + \Sigma^2}
\label{eq:sigma}
\end{equation} 
We find the solution $\Sigma \ne 0$, which breaks chiral symmetries, if
\begin{equation}
\frac {1}{g^2} < \frac {1}{g_c^2} = \frac {2 \Lambda}{\pi^2}
\end{equation} 
(where $\Lambda$ is the ultraviolet cutoff in the momentum integral above),
hence the phase diagram of Fig.\ref{fig:FIG5}

Consider now the effects of temperature or density, i.e.
we ideally add another axes to the phase diagram of 
Fig. \ref{fig:FIG5}. If we start increasing temperature and density 
from the symmetric phase, things  do not change much, 
while in the broken phase high temperature and/or density lead to the
restoration of the chiral symmetries.  This is intuitive if we 
just consider the disorder induced by temperature
(of course the spins will become disorded in chiral space,
not in the real one).
This behaviour can be revelaed by a mean field
analysis. Let us first  introduce explicitely temperature,
as discussed in Sect. 2.3, by replacing the intergral
over time by a sum over discrete Matsubara frequencies
\begin{equation}
\int d^3 p \rightarrow \sum_{Matsubara frequencies} \int d^2 p
\end{equation}
and then include a chemical potential $\mu$ for baryon number $N_b$ by
adding to the Lagrangean a term $\mu N_b$:
\begin{equation}
\mu N_B \rightarrow \mu J_0 = \mu \bar \psi \gamma_0 \psi
\end{equation}
(remember $J_0$ is the 0-th component of the conserved current
$\bar \psi \gamma_\mu \psi$). 
All in all, the finite temperature--finite density generalisation
of eq. 54 reads:
\begin{equation}
\frac{1}{g^2} = 4T \sum_{n = - \infty}^{n = + \infty} \int
\frac{d^2 p}{2 \pi ^2 } \frac{1} 
{((2n -1)\pi T - i \mu)^2 + p^2 + \Sigma^2(\mu, T)}
\end{equation}
We can now eliminate the coupling in favour of $\Sigma_0 =
\Sigma(T = 0, \mu = 0) $
\begin{equation}
\Sigma - \Sigma_0 = - T [ \ln ( 1 + e^{-(\Sigma - \mu)/T})
+ ln( 1 + e^{-(\Sigma + \mu)/T}]
\end{equation}
This gives the behaviour of the order parameter at fixed temperature 
as a function of $\mu$.
A chiral restoring phase transition is apparent for any temperature. At $T=0$ 
the transition is (very strongly) first order, and occurs
for $\mu_c = m_F (\mu=0)$, as expected of simple models. 
For any other temperature the
transition is second order, and $\mu_c$ gets smaller
and smaller while increasing the temperature, eventually becoming $0$
at $T = T_c$. A similar calculation gives the equation of state,
in particular for $T = 0$ the baryon number $N_b$ is seen to 
(approximatively) follow the prediction of a free 
(massless) fermion gas
$N_b \propto \mu^3$.
\begin{figure}[htb]
\begin{center}
{\epsfig{file=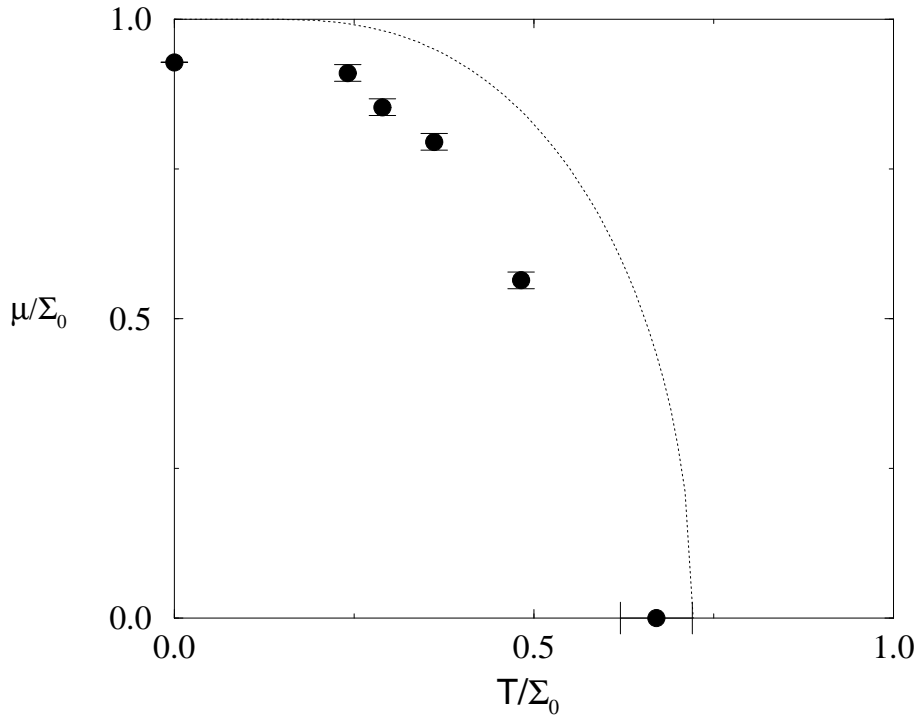, width=12 truecm}}
\caption{\footnotesize The phase diagram of the 3d Gross--Neveu model
in the $T,\mu$ plane. The coupling has been selected in such a way 
that the model is in the symmetry broken phase at $T=\mu=0$.
Chiral symmetry restoration at large $T,\mu$ is observed either
in mean field calculation (solid line) and exact lattice 
simulations (points). Taken from S. Hands, ref. \cite{hands_kogut}}
\end{center}
\label{fig:hands}
\end{figure}
The behaviour of the order parameter can be used to draw the
phase diagram in the temperature--chemical potential plane,
Figure \ref{fig:hands}

But what happens beyond mean field? Or, equivalently, what happens
when the number of flavor decreases from infinite to any finite
number? (the results of the mean field analysis can be shown to
be equivalent to the leading order of a $1/N_f$ expansion).
Although it is possible to calculate $1/N_F$ corrections
it is clearly desirable to have some exact results. 
For instance, one wants to find out exactly the position of
the tricritical point where the first order line merges
with the second order one (mean field predicts $\mu=0$).
Or, one might inquire about the  stability of the nuclear matter 
phase, i.e. for which range of temperature one can excite massive
fermions, before restoring the chiral symmetry?
These and other points have been studied in \cite{kos}:
the authors have shown that dimensional reduction is valid at 
the finite T transition, discussed the existence of 
a tricritical point and the possibility of a nuclear
liquid--gas transition.
Clearly these aspects depend on the exact dynamics. We have seen
in the discussion on the electroweak transition that mean field
can give completely misleading indications.
A feeling about the relevance of the quantum corrections in this
model  can be obtained by comparing
the phase diagram obtained on the lattice, and via the mean field
calculation described here (see Figure 7) : 
we note that there is no qualitative
change, however there are sizeable differences. 

Let us summarize: 
the phase diagram of four fermion models can be studied
within a  self consistent mean field approach, equivalent
to a leading order expansion in $1/N_f$. These models
are amenable to an exact lattice study (at variance with QCD 
whose lattice study, as we shall see,  is limited to $\mu=0$).
The results available on the lattice, while confirming many
qualitative trends, show also significant deviations from a
simple mean field analysis. Going from simple 
calculations of simple models to exact calculations 
of the same models can give interesting, 
new information, and much is still to be done in this  field. 

\section{The Phases of QCD}

Let us recall the symmetries of the QCD action with
$N_f$ flavors of massless quarks, coupled to a $SU(N_c)$ 
color group:
\begin{equation}
SU(N_c)_C \times SU(N_f) \times SU(N_f) \times Z_A(N_f)
\end{equation}
$SU(N_c)$ is the gauge color symmetry.
$SU(N_f) \times SU(N_f) \times Z_A(N_f)$ is
the flavor chiral symmetry, after the breaking of the classical
$U_A(1)$ symmetry to the discrete $Z_A(N_f)$.

We want to study the realisation and pattern(s) of breaking
of the chiral symmetries 
and we would like to know the interrelation of the above
with the possibility of quark liberation predicted 
at high temperature and density\cite{capa}.

The first task -- fate of
the chiral symmetries -- is made difficult by the problems with simulating
QCD with three colors at finite baryon density. Aside from this, the tools 
for investigating the transition are all at hand:
significant results at finite temperature 
have been obtained from first principle
calculations on the lattice, while the phase diagram
of the two color model (which can be simulated at nonzero
baryon density) is well underway. For a recent phenomenological
approach to  finite density QCD, not discussed here, 
we refer again to \cite{vari}. 
The second task -- confinement -- poses more conceptual problems.
It can be addressed satisfactorily only in Yang Mills systems,
which enjoy the global $Z(N_c)$ symmetry of the center of the
gauge group. When light quarks enter the game, the global $Z(N_c)$ 
symmetry is lost, and the simple description of confinement
in terms of such symmetry is not possible any more. 

In normal condition (zero temperature and density) the
$SU(N_f)_L\times SU(N_f)_R$ chiral symmetry is spontaneously broken
to the diagonal $SU(N_f)_{L+R}$. This should be signaled
by the appearance of a Goldstone boson, and a mass gap. 
In the real world, as such a symmetry
is only approximate,  the would be Goldstone is not exactlyt massless,
but it is neverthless lighter than any massive baryon --i.e.
there is a `relic' of the mass gap. All of them can be described
by  chiral perturbation theory, where the small expansion parameter
is, essentially, the bare quark mass. 

Most dramatic dynamical features of QCD are confinement and asymptotic
freedom: in ordinary condition quarks are `hidden' inside hadrons,
i.e. all the states realised in nature must be color singlet. 
If one tries to `pull' one quark and one antiquark infinitely apart, 
the force between them grows with distance--which is also called infreared
slavery. At short distance, instead, 
quarks are nearly free: this is called asymptotic freedom.

In summary,  these are the features of QCD we shall be concerned with
in the next two Sections:
\begin{itemize}
\item Asymptotic freedom
\item Confinement 
\item Spontaneous chiral symmetry breaking 
\end{itemize}
Confinement and relisation of chiral symmetry depend on the thermodynamic
conditions, and on the number of flavors and colors. Asymptotic freedom
does not depend on thermodynamics: it holds true till
\begin{equation}
N_f < 11/2 N_c
\end{equation}
\begin{figure}[htb]
\begin{center}
{\epsfig{file=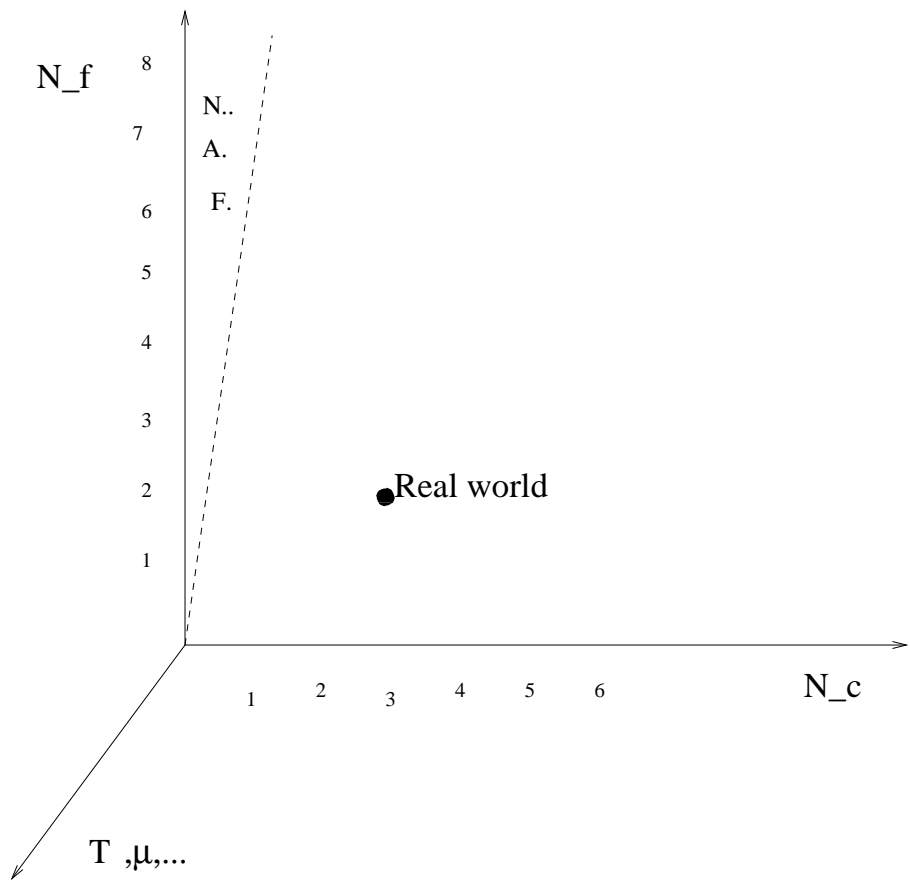, width=12 truecm}}
\caption{\footnotesize The phase diagram QCD in the $N_f$, $N_c$,
and thermodynamic parameters space. The everyday world corresponds to
$N_c=3$, $N_f=2$ (approximatively), $T=\mu=0$. For $T=\mu=0$ there is
a line in the $N_c$,$N_f$ plane separating the `ordinary' phase
of QCD from one without asymptotic freedom.}
\end{center}
\label{fig:FIG6}
\end{figure}
 For $T=\mu=0$ there is then 
a line (see Fig. \ref{fig:FIG6}) 
in the $N_c$,$N_f$ plane separating the `ordinary' phase
of QCD from one without asymptotic freedom.  More exotic phases
have been predicted as well \cite{appsann}, but it is 
anyway well established that it
does exhist a region in the $N_f,N_c$ plane which is characterised
by confinement, spontaneous chiral symmetry breaking and asymptotic
freedom.
The third axes can represent either temperature and density: one
might well imagine that the any point belonging in the same sector
of the phase diagram in  Fig. \ref{fig:FIG6} 
would behave in a similar way while increasing
temperature and density. It is then interesting to study 
the thermodynamics of  models
with different $N_c$ and $N_f$, looking for similarities and
differences.

In the discussion of the next two Sections 
we will consider $N_c=3$, the real world, 
and $N_c=2$, for technical reasons, and  because it represents an 
interesting limiting case. We will discuss  the
Yang Mills model, which also describes QCD with quarks of infinite mass.
We will mostly discuss two light flavors, which is rather realistic,
as the up and down quarks are much lighter than any other mass.
Understanding the behaviour
with three flavor is important as well, as in the end the real
world is somewhere in between $N_f=2$ and $N_f=3$.

\section {QCD at Finite Temperature,  $\mu_B = 0$}

Disorder increases with temperature. Then, one picture of the high $T$
QCD transition can be drawn by using the ferromagnetic analogy of
the chiral transition mentioned in Section 2: 
$\bar \psi \psi$ can be thought of as a spin
field taking values in real space, 
but whose orientation is in the chiral sphere. 
Chiral symmetry breaking occurs
when $< \bar \psi \psi> \ne 0 $, i.e. it corresponds to the ordered phase.
By increasing T, $< \bar \psi \psi> \to 0$. This is very much the same
as in the Gross Neveu model of Section 5. 
Color forces at large distance should decrease with temperatre: the main
mechanism, already at work at $T=0$, is the recombination
of an (heavy) quark and antiquark with pairs generated by the
vacuum: $\bar Q Q \rightarrow \bar q Q + q \bar Q$. 
At high temperature it becomes easier to produce light
$\bar q q$ pairs from the vacuum, 
hence it is easier to `break' the color string  
between an (heavy) quark  and antiquark $\bar Q Q$. 
In other words, we expect screening of the color forces.
It is however worth mentioning that, even if the string `breaks' 
bound states might well survive giving rise to a complicated, 
non--perturbative dynamics above the critical temperature.
The physical scale of these phenomena is  the larger physical
scale in the system, i.e. the pion radius (cfr. the discussions
in Section 2).

There are two important limits which are amenable to a
symmetry analysis : $m_q = 0$ and $m_q = \infty$.

\subsection{QCD  High T P.T., and Symmetries I : $m_q=0$ and
the Chiral
Transition}

When $m_q = 0$  the chiral symmetry of eq.(64) is exact.
As both  $m_u$ and $m_d$ are much smaller than $\Lambda_{QCD}$ , 
this is a reasonable approximation.
Note the isomorphy
\begin{equation}
SU(2) \times SU(2) \equiv O(4)
\end{equation}
which shows that the symmetry is the same as the one of an O(4) ferromagnet.
The relevant degrees of freedom are the three pions, and the sigma
particle, and the effective potential is a function of 
$\sigma^2 +  |\pi|^2$ in the chiral space. Once a direction in the
chiral sphere is selected (say in the $\sigma$ direction) chiral
symmetry is spontaneously broken in that direction, according
to the pattern:
equivalently
\begin{eqnarray}
 SU(2)_R \times SU(2)_L & \rightarrow & SU(2)_{L + R} \\
 O(4) & \rightarrow & O(3) 
\end{eqnarray}
Massless Goldstone particles (in this case, the three pions) 
appear in the direction ortogonal to the one selected by the
spontaneous breaking. 

Combining this symmetry analysis with the general idea of dimensional
reduction, Pisarski and Wilczek 
\cite{piwi} proposed that the high temperature transition
in the two flavor QCD should be in the universality class of the O(4)
sigma model in three dimensions.
At high temperature when symmetry is restored there will be just one global
minimum for zero value of the fields, and  pion and sigma become eventually 
degenerate.

We have however to keep in mind possible sources of violation of
this appealing scenario (see \cite{kosi} for recent discussions).

Firstly, is the very nature of the symmetry which is 
restored across the transition. \cite{piwi} \cite{shuwhich} 
Note infact that the $\sigma$ model picture
assumes that the axial $U(1)$ symmetry 
remains broken to $Z(N_f)$ across the
phase transition, in such a way that it does not impact on the
critical behaviour. This is indeed a dynamical question -- which symmetry
is restored first -- which can only be answered by an
ab initio calculation.

One second possible source of violation of this scenario relates
with a strong deconfining transition happening for its own reasons.
Such deconfining transition would liberate abrubtely 
an huge amount of degrees
of freedom, and   would likely trigger also the restoration of chiral symmetry.
In this case, the chiral transition
 would be hardly related with the sigma model.

And there are also practical considerations: 
maybe the scenario is true, but the
scaling window is so small that it has no practical applicability
or relevance. One example of this behaviour is offered by the
$Z(N_f) \times Z(N_f)$ 3 dimensional Gross-Neveu model. Its high
T phase transition should be  in the universality class of the Ising model,
but the scaling window turns out to be only $1/N_f$ wide, i.e.
it shrinks to zero in the large $N_f$ (mean field) limit, in which case
dimensional reduction is violated.


All in all, one has to resort to numerical simulations to measure
the critical exponents, and verify or disprove the $O(4)$ universality.
In turn, this gives information on the issues (nature of the chiral
transition, role of deconfinement, heavy modes decoupling, etc.)
raised above.

In practice, one measures the chiral condensate as a function of
the coupling parameter $\beta$, which in turns determines the temperature
of the system. This gives the exponent $\beta_{mag}$ (remember the discussions
in Section I) according to
\begin{equation}
<\bar \psi \psi> = B (\beta - \beta_c)^{\beta_{mag}}
\end{equation}
The exponent $\delta$ is extracted from the repsonse at criticality:
\begin{equation}
<\bar \psi \psi> = A m^{1/\delta} ; \beta = \beta_c
\end{equation}
The results (we quote only ref. 
\cite{kosi} as there is a substantial agreement) 
: $\beta_{mag} = .27(3)$, $\delta = 3.89(3)$
compare favourably with the O(4) results
$\beta_{mag} = .38(1)$, $\delta = 4.8(2)$,
and definitively rule out mean fields exponets (which would have characterised
a weak first order transition). However, the results can still be compatible
with O(2) exponents, which would signal the persistence of some lattice
artifact, $\beta_{mag} = .35(3)$, $\delta = 4.8(2)$, and of course it
is still possible that the final answer do not fit any of the above
predictions, for instance one is just observing some crossover
phenomenon.

\subsection{Two Color QCD I}

Two color QCD ejoys an enlarged (Pauli--G\"ursey) chiral symmetry:
quarks and antiquarks belong to equivalent representation of the
color group (see for instance the first entry of \cite {su2den},
and references therein) . 
As a consequence of that the ordinary chiral symmetry of QCD
\begin{equation}
SU(N_f)\times SU(N_f)
\end{equation}
is enlarged to $SU(2N_f)$. The spontanoues breaking of 
$SU(2N_f)$ then produces a different pattern of 
Goldstone bosons, which includes massles baryons (diquarks).
The relevant sigma model in this case includes also
diquarks and antidiquarks
Because of this symmetry, we have an exact degeneracy
of the pion, scalar $qq$ and scalar $\bar{q}\bar{q}$ and of the scalar meson,
pseudoscalar $qq$ and pseudoscalar $\bar{q}\bar{q}$ : this is precisley
the reason why we have massless baryons in the model, and thus why
this is not, a priori, a good guidance to QCD at finite baryon density
(to be discussed in Section 8).

The Pauli--G\"ursey symmetry 
has implications on the universality class of the chiral
transition at high temperature: for instance, the 
predictions of universality+dimensional reduction for the
two color, two flavour is $O(6)$ critical exponents
\cite{wirstam}.
The general argument  is clear, but it has
been fully appreciated and studied only recently. Also, this
tells us that qualitative arguments relating
the universality class of the chiral  transition to the 
flavour, but not to the color group,  have to be used with
care: at least for two color QCD, which is an interesting
limit case, the number of color matters for the high temperature
chiral transition as well.

\subsection{QCD High T p.t., and Symmetries II : $m_q = \infty$ 
and the Confinement Transition}
When $m_q = \infty$  quarks are static and  do not
contribute to the dynamics : hence, the dynamic of the system is 
driven by gluons alone, i.e. we are dealing with a purely Yang-Mills
model:
\begin{equation}
S = F_{\mu \nu}F^{\mu \nu}
\end{equation}
 In addition to the local gauge symmetry, the action enjoys the global 
symmetry associated with the center of the group, $Z(N_c)$. 
The order parameter is the Polyakov loop $P$
\begin{equation}
P = e ^ { i\int_0^{1/T} A_0 dt} 
\end{equation}
In practice, $P$ is the cost of a static source violating the $Z(N_c)$
global symmetry.

The interquark potential V(R,T) (R is the distance, T is the temperature)
is
\begin{equation}
e^{-V(R,T)/T} \propto < P(\vec 0) P^\dagger (\vec R) >
\end{equation}
Confinement can then be read off the behaviour of the interquark potential
at large distance.
When  $V(R) \propto \sigma R$  it would cost an infinite
amount of energy to pull two quarks infinitely apart. Above a certain
critical temperature $V(R)$ becomes constant at large distance: i.e.
the string tension is zero, confinement is lost. 
The implication of this is that $|P|^2 = V(\infty, T)$ 
is zero in the confining phase, different from zero otherwise.
P plays thus a double role, being the order parameter of the
center symmetry, and an indicator of confinement
(for an alternative symmetry description of the pure gauge deconfinement
transition see \cite{kov}). 
We learn that in Yang Mills models there is
a natural connection between confinement and realisation of the
$Z(N_c)$ symmetry. Hence,  the confinement / deconfinement
transition in Yang Mills systems is amenable to a symmetry
description.
By applying now the same dimensional reduction argument as above,
we conclude that the Universality class expected of
the three color model is the same as the one of a three
dimensional model with $Z(3)$ global symmetry: this is 
the three state Potts model discussed in Section 3.3.
Indeed, the transition turns 
out to be `almost' second order, i.e. very weakly first order,
like the 3d three state Potts model,
see \cite{pisa} for review and phenomenological implications
of this observation.

\subsection{Two Color QCD II}
The same reasoning tells us that the two color model is in the universality 
class of the three dimensional $Z(2)$ (Ising) model. 
This prediction has been checked with
a remarkable precision \cite{engels}, and it is a spectacular confirmation of
the general idea of universality and dimensional reduction. 
$$\vbox{\offinterlineskip
\halign{
\strut\vrule     \hfil $#$ \hfil  &
      \vrule # & \hfil $#$ \hfil  &
      \vrule # & \hfil $#$ \hfil  &
      \vrule # & \hfil $#$ \hfil
      \vrule \cr
\noalign{\hrule}
  ~{\rm Source}~
&&~~
&&~{SU(2)}~
&&~{\rm Ising}~\cr
\noalign{\hrule}
  ~~<|L|>~~&&~~\beta/\nu~~&&~~0.525(8)~~~&&~~0.518(7)~~~~\cr
  ~D<|L|>~~&&~~(1-\beta)/\nu~~&&~~1.085(14)~~&&~~1.072(7)~~~~\cr
 ~~~~~~~~~&&~~1/\nu~~&&~~1.610(16)~~&&~~1.590(2)~~~~\cr
 ~~~~~~~~~&&~~\nu~~&&~~0.621(6)~~~&&~~0.6289(8)~~\cr
 ~~~~~~~~~&&~~\beta~~&&~~0.326(8)~~~&&~~~0.3258(44)~\cr
\noalign{\hrule}
 ~~~~\chi_\nu~~~&&~~\gamma/\nu~~&&~~1.944(13)~~&&~~1.970(11)~~~\cr
 ~~~D\chi_\nu~~~&&~~(1+\gamma)/\nu~~&&~~3.555(15)~~&&~~3.560(11)~~~\cr
 ~~~~~~~~~&&~~1/\nu~~&&~~1.611(20)~~&&~~1.590(2)~~~~\cr
 ~~~~~~~~~&&~~\nu~~&&~~0.621(8)~~~&&~~0.6289(8)~~\cr
 ~~~~~~~~~&&~~\gamma~~&&~~1.207(24)~~&&~~1.239(7)~~~~\cr
\noalign{\hrule}
 ~~~~~~~~~&&~~\gamma/\nu+2\beta/\nu~~&&~~2.994(21)~~&&~~3.006(18)~~\cr
\noalign{\hrule}
 ~~~~g_r~~&&~~-g_r^{\infty}~~&&~~1.403(16)~~&&~1.41~~~~~~~\cr
 ~~~Dg_r~~&&~~1/\nu~~&&~~1.587(27)~~&&~~1.590(2)~~\cr
 ~~(\omega=1)~~&&~~\nu~~&&~~0.630(11)~~&&~~0.6289(8)~\cr
\noalign{\hrule}}
}$$
\vskip 5pt

\centerline { Taken from 
J.~Engels, S.~Mashkevich, T.~Scheideler and G.~Zinovjev, 
ref. \cite{engels} }

\subsection{Summary and Open Questions for the QCD High T Transition}

In Figure 7 we sketch a phase diagram in the temperature, bare mass
plane for a generic version of QCD with $N_f$ flavor degenerate in mass,
and $N_c$ color. 

For zero bare mass the phase transition is chiral.
For three colors, two flavors is second order with $T_c
\simeq 170 Mev$. The prediction from dimensional reduction + universality
--$O(4)$ exponenents-- 
is compatible with the data, but the agreement is not perfect.
If the agreement were confirmed, that would be an argument in favour
of the non-restoration of the $U_A(1)$ symmetry at the transition,
which is also suggested by the behaviour of the masses spectrum.
Remember infact that the chiral partner of the pion is
the $f_0$, which is in turn degenerate with the scalar
$a_0$ with $U_A(1)$ is realised. All in all, 
$U_A(1)$ non--restoration across the chiral transition 
corresponds to $m_\pi \simeq m_{f_0} 
\ne m_{a_0}$ which is the pattern observed in lattice calculations
\cite{karsch}.
The transition with three (massless) flavour  turns out to be first
order. The question is than as to whether the strange quark should be 
considered `light' or heavy'. In general, the real world will be somewhere
in between two and three light flavour, and to really investigate
the nature of the physical phase transition in QCD one should work
as close as possible to the realistic value of the quark masses.

By switching on the mass term chiral symmetry is explitely broken
by $m < \bar \psi \psi>$. 
In the infinite mass limit QCD reduces  to the pure gauge
(Yang Mills) model. Yang Mills systems have
a deconfining transition  associated with the realisation of the global 
$Z(N_c)$ symmetry.
This places the system in the Ising 3d universality class for two colors,
and makes the transition weakly first order (near second, infact) for
three colors. General universality arguments are 
perfectly fulfilled by the deconfining transition.

The $Z(N_c)$ symmetry is broken by the kinetic term 
of the action when the quarks are dynamic ($m_q < \infty$) :
this particular symmetry description of deconfinement
only holds for infinite quark mass. Till very recently then
the most convincing signals for deconfinement with dynamical
quarks come from the equation of state. For instance
the behaviour of the internal energy is a direct probe of the
number of degeres of freedom, and indicates quark and gluon liberation
\cite{karsch}.

Another important set of information come from the behaviour
of the mass spectrum--this is of course every relevant
both on experimental grounds, for the ungoing RHIC experiments
as well as for the upcoming ones, as well as for completing
our understanding of patterns of chiral symmetry. 
The most dramatic phenomena , i.e 
the disappearance of the Goldstone mode in the symmetric
phase, the nature of the Goldstone mechanism
in four fermion models, as well as the massive fermion in 
the broken one, have
been already studied on the lattice. 

Among the most prominent open questions, there is of course the
behaviour of `real' QCD, with two light flavour, and a third one
of the order of $\Lambda_{QCD}$, so how and when exactly the 
$N_f=2$ scenario morfs with the $N_f = 3$? Also, why is
$T_\chi$ much smaller that the pure gauge deconfining transition?
At a theoretical level the question is if it is possible to give
an unified description of the two transitions,
chiral and deconfining. This question is
currently under active investigation:
 recent work suggests that a symmetry analysis of the deconfining
transition can be extended also to theories with dynamical fermions.
The physical argument is rooted in a duality transformation
which allows the identification of magnetic monopoles as
agent of deconfinement. The order parameter for deconfimenent
would that be the monopole condensate \cite{digiaco}.
An alternative approach uses percolation as the common 
agent driving chiral and confining transitions
\cite{satz}.

\begin{figure}[htb]
\begin{center}
{\epsfig{file=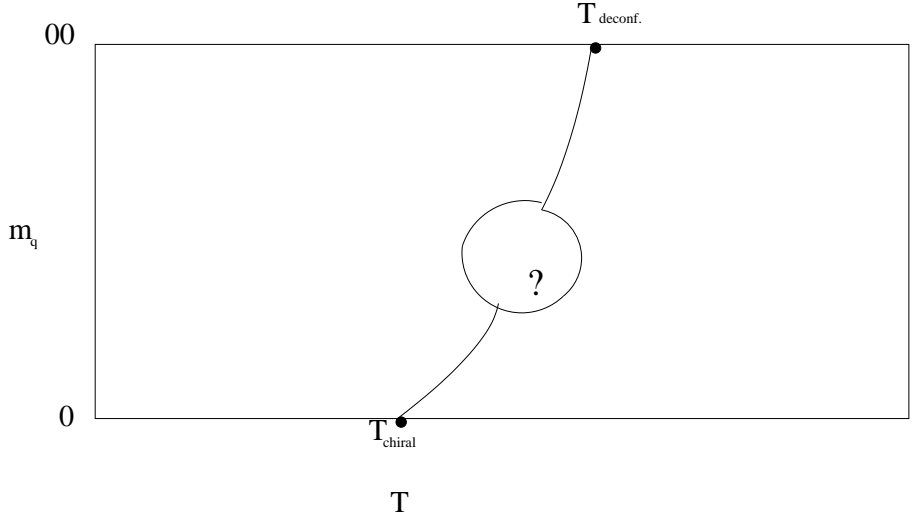, width=12 truecm}}
\caption{\footnotesize The `phase diagram' QCD in the quark mass,
temperature plane--see text.}
\end{center}
\label{fig:FIG7}
\end{figure}

\section{QCD at Finite Temperature,  $\mu_B \ne 0$}
Let us now consider the effects induced by an increased baryon density,
when baryons start overlapping. The most natural prediction relates with
asymptotic freedom: as the quarks get nearer and nearer they do not feel
the interactions any more, whilst the long range interaction is
screened by many body effects. The natural conclusion is that the system
is `nearly free' (i.e. non confining) hence
chiral symmetry is realised (one needs some interaction 
to break the symmetry of the action).
Both on physical grounds,
and from the predictions of simple models (Gross--Neveu, Section 5)
we expect the typical scale of critical phenomena at finite density 
to be  set by the lighter particle carrying baryon number.

First principle calculations should be able
to confront these predictions, as well as 
results obtained by use of semi-approximate
calculations of symmetry motivated models \cite{vari} and to put them on 
firm quantitative grounds. 

We address here three main points:
\begin{itemize}
\item Why is QCD at finite density difficult
\item What do we know from first principles
\item What can we do, in practice
\end{itemize}

The critical region of QCD is outside the reach of perturbative 
calculations, then one should use ab initio lattice methods. 
As most people have already heard, this is 
plagued by several problems. To understand the problems, and to propose
some solution, or, at least, workaround, we have now to give some 
(quick) detail of these 
calculations.

Consider again the partition function:
\begin{equation}
{\cal Z}(\mu, T) = \int_{0}^{1/T} dt \int e^{S(\bar \psi, \psi, U)} 
d \bar \psi d \psi dU
\end{equation}
A chemical potential for baryons is introducted by adding the term 
$\mu J_0$, where $J_0 = \bar \psi \gamma_0 \psi$ is the $0^{th}$
component of the conserved particle number current. In momentum
space, this corresponds to the substitution 
$p_0 \rightarrow p_0 - i \mu$.

Let us now specialise to the QCD Action
\begin{equation}
S_{QCD} = F_{\mu \nu}F_{\mu \nu} + 
\bar \psi (\not \partial + m + \mu \gamma_0) \psi =
S_G + \bar \psi M \psi
\end{equation}
and exploit the bilinear form of the fermionic part of the Action
to integrate explicitely over the fermions. This yields:
\begin{equation}
{\cal Z}(T, \mu) = \int dU det M e ^{-S_G} = \int dU e ^{-(S_g - log(det M))}
\end{equation}
The partition function can be {\em exactly}
written in terms of an Action which only contains gauge field,
$S_{eff} = S_G - log(det M)$ (we often call it an effective Action,
however we should remember that no information has been lost in
its definition)

Once the model is formulated on a lattice, one can express the
integral over space as a sum over the lattice sites, and the
relevant observables can be calculated by use of the importance
sampling. For this to be viable, $S_{eff}$ should be real.
The problem is that for three colors QCD $S_{eff}$ is not real.

The usual numerical methods then cannot be applied. Alternative methods
which have been proposed so far fail at zero temperature, but could
be applied, maybe, at nonzero temperature \cite{munonzerot}. 
Interesting new idea have been proposed, and tested in spin models
\cite{mureview}. All this is promising and gives some hope
that at least the region around $T_c$, small chemical potential can be
understood in the near future, see \cite{mureview} for a review
on recent efforts.

There is however something else one can try within the lattice formalism,
which is the subject of the next subsection.

\subsection{The Lattice Strong Coupling Analysis of the QCD Phase Diagram}

The lattice formulation, besides being suited for statistical MonteCarlo
simulations, lends also itself to an elegant expansion in terms of the
inverse gauge coupling (\cite {strong} \cite{hami}). 
As the continuum limit (i.e. real QCD) corresponds
to zero gauge coupling, at a first sight one might think that the
so-called strong coupling expansion is hopeless. But there is a very
important point: for two or three color QCD there are no phase transitions
as a function of coupling. Hence, many qualitative features - most
important, confinement and chiral symmetry breaking -- do not depend
on the gauge coupling itself. Moreover, this expansion is systematically
improvable: it is worth noticing a nice work where the 
asymptotic scaling associated with the continuum limit
was reached from strong coupling \cite{asystro}!!

Another interesting feature is that the strong coupling expansion
yields a purely fermionic model, with four fermion interactions. 
Then, the strong coupling expansion  can also be seen as a tool for 
deriving effective models for QCD from an ab inito formulation.

The starting point is the QCD lattice Lagrangean:
\begin{eqnarray}
S &=& -1/2 \sum_{x} \sum_{j=1}^3 
 \eta_j(x) [ \bar \chi(x) U_j(x) \chi(x + j) - \bar \chi (x + j)
U^\dagger_j(x) \chi (x)]  \\
&& -1/2 \sum_x \eta_0(x) [ \bar \chi(x) U_0(x) \chi(x + 0) 
- \bar \chi (x + 0) U^\dagger_0(x) \chi (x)]  \nonumber \\
&& -1/3 \sum_{x} 6/g^2 \sum_{\mu,\nu=1}^4[ 1 - re Tr U_{\mu \nu}(x)] 
\nonumber \\
&& + \sum_x m \bar \chi \chi \nonumber
\end{eqnarray}

The $\chi, \bar \chi$ are the staggered fermion fields living on the
lattice sites, the U's are the $SU(N_c)$ gauge connections on the links, the
$\eta$'s are the lattice Kogut--Susskind
counerparts of the Dirac matrices, and the chemical
potential is introduced via the time link terms $e^\mu$, $e^{-\mu}$ which
favour (disfavour) propagation in the forward (backward) direction
thus leading to the desired baryon-antibaryon asymmetry.

We have written down explicitely the lattice Action to show that the
pure gauge term 
$ S_G = -1/3 \sum_{x} 6/g^2 \sum_{\mu,\nu=1^4 }[ 1 - re Tr U_{\mu \nu}(x)]$
contain the gauge coupling in the denominator, hence it disappears in
the infinite coupling limit. Consequently, one can perform independent
spatial link integration, leading to
\begin{equation}
{\cal Z} = \int \prod_{time links} dU_t 
d \bar \chi d \chi e ^{-1/{4N} \sum_{ <x,y>}
\bar \chi (x) \chi (x) \bar \chi (y) \chi (y) }
e^{-S_t}
\end{equation}
where $\sum_{<x,y>}$ means sum over nearest neighbooring links, terms
of higher order have been dropped, and we recognize a four fermion
interaction. From there on things proceed formally much in the same way
as for the Gross-Neveu model discussed in Section 5,
and we quote the final result:
\begin{equation}
{\cal Z} = \int Z_0^V d <\bar \chi \chi>
\end{equation}
where $V$ is the three dimensional space volume and $Z_0 =
 Z_0 (<\bar \chi \chi>, T, \mu)$   is the
effective partition function for $<\bar \chi \chi>$.
A standard saddle point analysis gives the condensate as a function
of temperature and density, and allows for the reconstruction of the
phase diagram. 

The results are the following : below a critical temperature, there
is a chiral transition as a function of chemical potential,
 closely correlated with a deconfining transition,
from a normal phase to a `quark--gluon plasma' phase. This transition
is first order, very strong at zero temperature, 
and weakens with temperature,
becoming second order at $\mu = 0, T = T_c$.
The phase diagram is qualitatively similar to that of the Gross
Neveu model, but for the fact that here we have also deconfinement
correlated with the chiral transition.
\subsection{The Phase Diagram of Two Color QCD}

At the beginning of this section it is useful to remind ourselves 
of the discussion in Section 7.2: in a nutshell, 
$SU(2)$ baryons are completely degenerate with mesons, 
hence there are massless baryons. 
Both on physical grounds,
and from the predictions of simple models (Gros--Neveu, Section 5)
we know that the scale of critical phenomena at finite density is
set by the lighter particle carrying baryon number--the
massless diquark in two color QCD, and the massive baryon in
QCD. Hence,  two colors QCD  is not a good approximation to real QCD
at finite density.

Still, one might hope to learn something there, in particular 
concerning the gauge field dynamics and screening properties
at nonzero baryon density \cite{su2den}.
\begin{figure}[t]
\epsfxsize=20pc %
\epsfbox{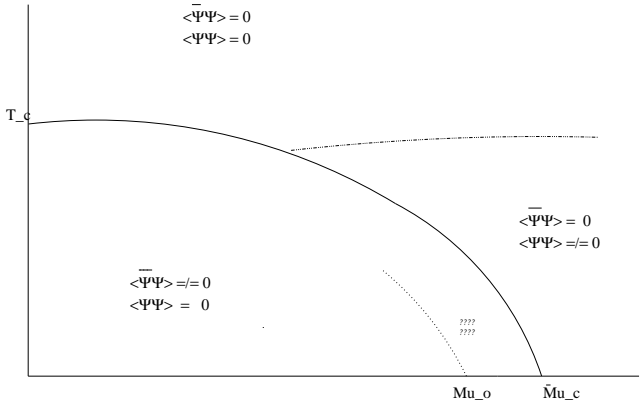} 
\caption{A sketchy view of the phase diagram of two colors QCD for
a non--zero quark mass}
\label{fig:su2pha}
\end{figure}
In Fig. \ref {fig:su2pha}  I sketch a possible 
(i.e. suggested by the results of
refs. \cite {su2den} )  
phase diagram of two color QCD,
in the chemical potential--temperature plane, for
an arbitrary, non--zero bare quark mass.
The general features of the theory at zero density , i.e. the
$\mu=0$ axis has been discussed in Sections 7.2 and 7.4 above.

When $\mu \ne 0$ quarks and antiquarks do not 
belong any more to equivalent representations: the Pauli--G\"ursey
symmetry is broken by a finite density.
When $\mu \ne 0$ the symmetry is reduced to that of `normal' QCD--
just because the extra symmetry quark-antiquark is explicitely broken:
the condensate will
tend to rotate in chiral space as $\mu$ increases, 
rotating into a purely diquark direction for large $\mu$.  
However, as $\mu$ increases and the symmetry in chiral space is reduced,
and the new vacuum would be physically
distinct from the original one. 
According to the standard scenario, at zero temperature,
for a chemical potential $\mu_o$
comparable with the mass of the lightest baryons, 
 baryons start to be produced 
thus originating a phase of cold, dense matter. 
For $SU(2)$ baryons (diquarks)  are bosons 
(as opposed to the fermionic  baryons of real
QCD). Besides the mass scale of the phenomenon,
there are then other important differences
between the dense phase in $SU(2)$ and $SU(3)$,
as, obviously, the thermodynamics of interacting
Bose and Fermi gases is different. In particular, diquarks
might well condense, however partial quark liberation is possible
as well.   We might then expect a
rather complicated ``mixed'' nuclear matter phase, perhaps
characterised by both types of condensates -- this is the region
marked by question marks in Figure 10.

To obtain a more direct probe of deconfinement, we can look at
the interquark potential by calculating the correlations 
$ <P(O) P^\dagger (z)>$ of
the zero momentum Polyakov loops, averaged over spatial directions.
This quantity is related to the string tension $\sigma$ via
$ <P(O) P^\dagger(z)> \propto e^{-\sigma z}$. Some of the results
of \cite{su2den} suggests indeed fermion screning, 
enhanced string breaking and the transition to a deconfined
phase.
Gauge field dynamics can also be probed by measurements of the
topological charge, as well as by the direct analysis of the
Dirac spectrum \cite{su2den}.

\subsection{The Phase Diagram of QCD}

We reproduce in Figure 11 the phase diagram of QCD from 
Ref. \cite{criqcd}.

Let us discuss the case of two massless flavors first.
The transition with $\mu = 0$ has been discussed at lenght 
in Section 7.5 above : it should be second order, in the
universality class of the O(4) model. The study of 
the high baryon density transition at $T=0$ meets with the problems discussed 
before in this Section. Neverthless, the model calculations
of \cite{vari}, the lattice numerical results for the Gross-
Neveu model reviewed in Section 5, the lattice strong coupling
results discussed above suggests a strong first order transition. 
If this is the case, a tricritical point should be found where the first order
line and the second order one merges along the phase diagram.
This is the point P. 

What happens if we consider light 
(rather than massless) up and down quarks, while still
keeping the strange very large? The possibility considered 
in \cite{criqcd} is that the second order line turns into a smooth
crossover, while the first order line is robust with respect to
this perturbation. The tricritical point now becomes the end point
E of the first order transition line. The effects of a not-so-heavy
strange quark mass can be very important : remember that the transition
for three massless flavors would be first order, hence for three
light flavor should remain first order as well, and the
point P would disappear. However this does not seem likely 
as the mass difference between up/down and strange is sizeable.

These aspects can be studied by considering the associate 
Ginzburg Landau
effective potential for the order parameter $<\bar \psi \psi>$,
and the critical properties of the point E can again be inferred by 
universality arguments. The same model calculations have produced 
estimates for the
position of the points P and E: $T_P\sim 100$ MeV
and $\mu_P\sim 600-700$ MeV. Of course these are
crude estimates as only chiral symmetry and not the full
gauge dynamics was taken into account.

Most important is the observation that the 
critical behaviour at the point E can be observed in ongoing
and future experiments at RHIC and LHC : indeed, E would
be a genuine critical point, even with massive quarks,
characterised by a diverging correlation length and 
fluctuations.

This is a very interesting situation where 
theoretical analysis
inspired by universality arguments, lattice
results at zero density, model calculations at zero temperature
have lead to a coherent picture and new predictions amenable
to an experimental verifications.

What next? Imaginary chemical potential calculations might 
perhaps be used to gain information around $T_c$ and small density: 
this might help refining our knowledge of points P and E. Moreover,
it might well be worthwhile to extend the lattice strong coupling
expansion by 1. including further terms 2. performing a more 
refined analysis  of the properties of the  effective potential. 
In particular it would be possible to study the position of
P, whose leading order estimate  is at $\mu = 0$:
the transition is always first order, with the only exception
of the zero density, high temperature transition. 
It would be very interesting to
see if the tricritical point would indeed move
inside the phase diagram. One final remark on
calculational schemes:  in addition to
the Lagrangean approach there is also the Hmiltonian one\cite{hami}.
Time is not discretised in the Hamiltonian approach, 
and for several reasons this might seem
more appropriate for dealing with a finite density. Moreoever, the
remarkable progress made by the condensed matter community on 
the sign problem over the last few years might perhaps be helpful
in this approach.

The scenario envisaged by \cite{digiaco}
\cite{satz} might have an impact  on the phenomenologial
issues considered here. Infact, if there is indeed one
critical line extending from the $m_q = 0$  axis
to $m_q = \infty$ as sketched in Fig. 8, the fate of
the chiral transitions with a finite quark mass should
be reconsidered. For instance, the second order chiral transition
with two  massless quarks would flow into the weak first order
transition of the pure Yang Mills model in the infinite
mass limit. Hence, apparently a small quark mass would 
not wash out the second order phase transition.
Once again, only ab initio lattice calculations can settle
the issue.

\begin{figure}[htb]
{\epsfig{file=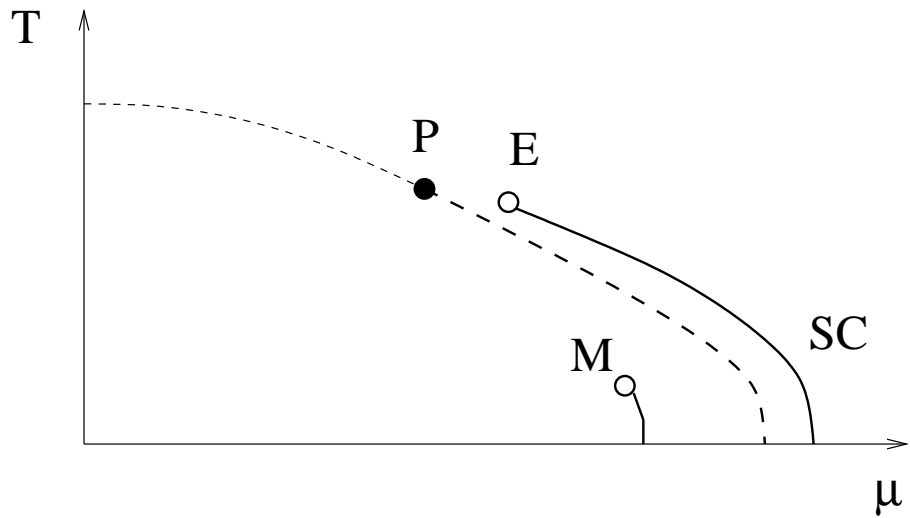,width=12 truecm}}
\caption[]{The schematic phase diagram of QCD, from
ref. \cite{criqcd}. The dashed line is for the chiral
transition for two flavor QCD, which is second order at 
$\mu=0$ and most likely first order at T=0, hence a
tricritical point P in between. The solid line is
the relic of such phase transition with a small quark mass,
and the tricritical point P turns into the endpoint E. 
The nuclear matter phase of QCD starts beyond the short line
ending by M, which can be experimentally studied. The authors
of \cite{criqcd} \cite{criexp} have proposed experimental
signatures to demonstrate the exhistence of,
and locate the point E }
\label{fig:steshura}
\end{figure}

\section*{Acknowledgments}
I would like to thank the Organisers for a most enjoyable time
in Trieste, as well as the Participants for many interesting
conversations.  In addition I thank the 
Institute for Nuclear Theory at the University of Washington
and the U.S. Department of Energy for partial support.
Part of the work described here was carried out at the Gran
Sasso Laboratory in Italy, and while visiting  the European
Center for Theoretical Studies in Nuclear Physics and Related
Areas, ECT$^*$, Trento, Italy.


\begin{thebibliography}{99}
\bibitem{chiral_news} There has been steady progress on 
 chiral fermions, the field is evolving, several formulations 
are currently under scrutiny,
and are already producing results, which will hopefully soon
reach the same quantitative accuracy of  standard calculations.
I would like to thank  S. Randjbar--Daemi for 
conversations on this subjects.
\end{thebibliography}
\end{document}